\documentclass[twocolumn,aps,pra,superscriptaddress,longbibliography]{revtex4-1}
\usepackage{color}
\usepackage{amsmath}
\usepackage{amssymb}
\usepackage{graphicx}
\usepackage{tikz}
\usepackage{multirow}
\usepackage{float}
\usepackage{bbm}
\usepackage{dsfont}
\usepackage{cases}
\usepackage{mathtools}

\usepackage{epsfig}
\usepackage{verbatim}
\usepackage{array}
\usepackage{setspace}

\usepackage{scalerel}
\usepackage{stackengine,wasysym}
\usepackage{empheq, nccmath}

\usepackage[T1]{fontenc}

\usepackage[unicode=true,
bookmarks=true,bookmarksnumbered=false,bookmarksopen=false,
breaklinks=false,pdfborder={0 0 1},backref=false,colorlinks=true]
{hyperref}
\hypersetup{
	linkcolor=magenta, urlcolor=blue, citecolor=blue, pdfstartview={FitH}, hyperfootnotes=false, unicode=true}

\makeatletter
\@ifundefined{textcolor}{}
{%
	\definecolor{BLACK}{gray}{0}
	\definecolor{WHITE}{gray}{1}
	\definecolor{RED}{rgb}{1,0,0}
	\definecolor{GREEN}{rgb}{0,1,0}
	\definecolor{BLUE}{rgb}{0,0,1}
	\definecolor{CYAN}{cmyk}{1,0,0,0}
	\definecolor{MAGENTA}{cmyk}{0,1,0,0}
	\definecolor{YELLOW}{cmyk}{0,0,1,0}
}

\begin{document}

\preprint{APS/123-QED}

\title{Free-fermion approach to the partition function zeros : Special boundary conditions and product form of solution}

\author{De-Zhang Li}
\address{Quantum Science Center of Guangdong-Hong Kong-Macao Greater Bay Area, Shenzhen 518045, China}

\author{Xin Wang}
\thanks{Corresponding author: x.wang@cityu.edu.hk}
\address {Department of Physics, City University of Hong Kong, Hong Kong SAR, China}
\affiliation{City University of Hong Kong Shenzhen Research Institute, Shenzhen 518057, China}

\begin{abstract}
Partition function zeros are powerful tools in understanding critical behavior. In this paper we present new results of the Fisher zeros of two-dimensional Ising models, in the framework of free-fermion eight-vertex model. First we succeed in finding special boundary conditions for the free-fermion model, under which the partition function of a finite lattice can be expressed in a double product form. Using appropriate mappings, these boundary conditions are transformed into the corresponding versions of the square, triangular and honeycomb lattice Ising models. Each Ising model is studied in the cases of a zero field and of an imaginary field $i(\pi/2)k_BT$. For the square lattice model we rediscover the famous Brascamp-Kunz (B-K) boundary conditions. For the triangular and honeycomb lattice models we obtain the B-K type boundary conditions, and the Fisher zeros are conveniently solved from the product form of partition function. The advantage of B-K type boundary conditions is that the Fisher zeros of any finite lattice exactly lie on certain loci, and the accumulation points of zeros can be easily determined in the thermodynamic limit. Our finding and method would be very helpful in studying the partition function zeros of vertex and Ising models. 
\end{abstract}

\maketitle

\section{Introduction}   \label{intro}
Since the seminal work of Lee and Yang in 1952 \cite{RN303, RN57}, the method of partition function zeros has played a crucial role in the study of critical phenomena. The Lee-Yang formalism provides a rigorous framework connecting the partition function zeros to phase transitions in general systems \cite{RN479}. In the context of Ising models, the so-called Lee-Yang zeros are typically defined in the complex magnetic-field plane. Lee and Yang also proved the celebrated circle theorem \cite{RN57}, which characterizes the distribution of these zeros for the case of ferromagnetic interactions. Fisher later extended Lee and Yang's idea to analyze the singularities in the complex temperature plane \cite{RN308}; the corresponding zeros are now known as Fisher zeros.

In this work, we focus on the Fisher zeros of two-dimensional Ising models $\left\{ {s_i} = \pm 1,~i = 1, \cdots, N \right\}$ with nearest-neighbor interactions. The Hamiltonian takes the form
\begin{equation}
H = J\sum\limits_{\left\langle {ij} \right\rangle} {s_i}{s_j} - H_{\rm{ex}}\sum\limits_i {s_i},     \label{eq1}
\end{equation}
where $J$ is the interaction constant and the sum $\sum_{\left\langle {ij} \right\rangle}$ is taken over all nearest-neighbors. Since in most exactly solvable cases the field $H_{\rm{ex}}$ is either 0 or $i(\pi/2)k_BT$ \cite{RN72, RN57, RN122, RN123, RN81, RN121, RN82, RN67, RN68, RN58, RN51, RN274, RN558}, we only consider these two values for $H_{\rm{ex}}$.

For the square lattice Ising model in the zero field, Fisher demonstrated that the partition function zeros in the complex $z=e^{2\beta J}$ plane lie on two circles $\pm 1 + \sqrt 2 e^{i\theta}\left(0 \le \theta < 2\pi \right)$ in the thermodynamic limit \cite{RN308}. He utilized the well-known Onsager–Kaufman solution of the finite lattice partition function under the toroidal (periodic in both directions) boundary conditions (BCs) \cite{RN73}. However, under these BCs, the partition function consists of the sum of four product terms, making the exact calculation of Fisher zeros of a finite lattice analytically intractable. Numerical calculations of Fisher zeros have since been conducted \cite{RN527, RN528, RN505, RN537, RN506, RN428}, and many studies of the partition function under various alternative BCs have been reported \cite{RN459, RN410, RN513, RN477, RN522, RN432, RN455, RN465, RN475, RN508, RN509, RN514, RN512}. A breakthrough came with the introduction of the Brascamp-Kunz (B-K) BCs \cite{RN410}, under which the partition function of a finite square lattice in the zero field can be given in a double product form. The use of B-K BCs enables explicit calculation of the Fisher zeros of a finite lattice, which precisely lie on the two circles in the complex $z=e^{2\beta J}$ plane, thus providing a rigorous determination of Fisher loci in the thermodynamic limit. The B-K BCs had been extended to the case of the imaginary field $H_{\rm{ex}}=i(\pi/2)k_BT$, where the finite lattice partition function also takes a double product form and the Fisher zeros remain confined to well-defined loci \cite{RN432}. Lu and Wu made use of the advantage of B-K BCs, to derive the density distribution function of Fisher zeros in the thermodynamic limit \cite{RN411}. A knowledge of density function of Fisher zeros can provide valuable insights into the physics of the system, e.g., phase transition and low-temperature series expansion \cite{RN529}. 

\begin{figure*} 
\includegraphics{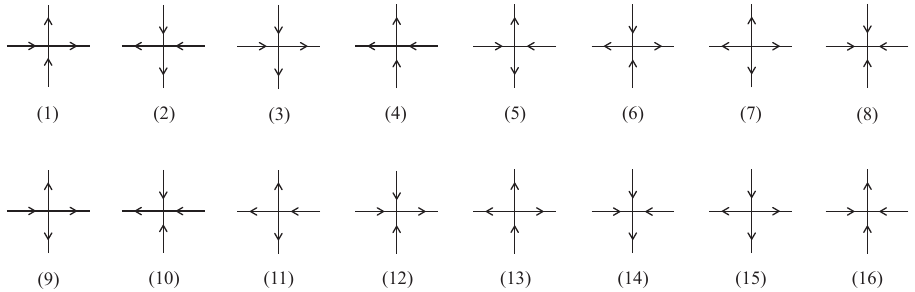}
\caption{The vertex configurations of the sixteen-vertex model. The even eight-vertex model consists of vertices (1)--(8), while the odd subcase consists of vertices (9)--(16).} \label{fig1}
\end{figure*}

In this work, we contribute to the picture by extending the B-K type BCs to the triangular and honeycomb lattice Ising models. Our approach is to formulate the Ising problem within the framework of the free-fermion eight-vertex model \cite{RN65, RN59, RN129}. This idea is directly inspired by a systematic study of the sixteen-vertex model presented in Ref. \cite{RN124}. As shown in Fig. \ref{fig1}, the sixteen-vertex model on the square lattice includes two subsets: the even eight-vertex configurations [vertices (1)–(8)] and the odd ones [vertices (9)–(16)]. In Ref. \cite{RN124} the author attempted to find suitable BCs for the even and odd eight-vertex models such that the finite lattice partition function has a simple product form, but did not succeed. In this study we have identified such BCs for the eight-vertex model satisfying the free-fermion condition, in both even and odd subcases. Using the mapping method developed in Ref. \cite{RN65}, we translate these BCs into the corresponding versions of the Ising models on the square, triangular and honeycomb lattices. As we will demonstrate later, this leads to the B-K type BCs of the Ising models, under which the finite lattice partition function admits a double product form.

The remainder of this paper is organized as follows. In Sec. \ref{f-f}, we derive the partition function of the free-fermion model under the special BCs we find. We obtain exact double product forms in both the uniform and row-by-row staggered cases. Section \ref{squ} applies these BCs to rediscover the B-K BCs for the square lattice Ising model. In Secs. \ref{tri} and \ref{hon}, we further propose the B-K type BCs for the triangular and honeycomb lattice models, respectively. For each Ising model, both solutions in the zero field and in the imaginary field are presented. Summary and discussion are given in Sec. \ref{summary}.

\section{Special boundary conditions of the free-fermion model}   \label{f-f}
Given the weights $\left\{\omega _i,~i = 1, \cdots ,16 \right\}$ of the vertex configurations in Fig. \ref{fig1}, the well-known free-fermion condition is either 
\begin{equation}
\omega _1\omega _2 + \omega _3\omega _4 = \omega _5\omega _6 + \omega _7\omega _8   \label{eq2}
\end{equation}
for the even eight-vertex model \cite{RN65, RN59}, or
\begin{equation}
\omega _9\omega _{10} + \omega _{11}\omega _{12} = \omega _{13}\omega _{14} + \omega _{15}\omega _{16}   \label{eq3}
\end{equation}
for the odd eight-vertex model \cite{RN129}. 
Both the even and odd free-fermion models are exactly solvable. Among the various approaches developed to obtain the exact solution \cite{RN65, RN59, RN314, RN315, RN316, RN317, RN318, RN319, RN129, RN262}, we adopt the Pfaffian method to derive the partition function under the special BCs. The Pfaffian method has proven very useful in solving statistical lattice models, in particular the dimer model \cite{RN136, RN137, RN139, RN397, RN218, RN140, RN143, RN486, RN145, RN144, RN382, RN383, RN465}. When applied to the vertex models, this method usually requires the construction of a corresponding dimer lattice with appropriately assigned dimer weights. In our work, we follow closely the construction established in Ref. \cite{RN59}, which first proposed the Pfaffian solution for the even free-fermion model, but we implement a different choice of BCs.

Since Secs. \ref{squ}--\ref{hon} rely on the mapping from the Ising models to the even free-fermion model \cite{RN65}, we restrict our analysis in this section to the even subcase. All results presented in the main text pertain exclusively to the even subcase; the results of the odd subcase are provided in Appendix \ref{appa}. For clarity, we henceforth use the term "free-fermion model" to refer specifically to the even subcase throughout the main text. Additional discussions on the odd free-fermion model can be found in Appendix D.1 of Ref. \cite{RN124}.

\subsection{Uniform case}   \label{f-f-a}
Consider a square lattice with $M$ rows and $2N$ columns. We examine two sets of BCs, both of which impose periodic boundary conditions in the $N$ direction and prescribe fixed arrow configurations on the upper and lower edges of the lattice. These BCs are denoted as (\romannumeral1) and (\romannumeral2), corresponding to the following fixed-edge configurations:

\noindent (\romannumeral1). Both the upper edge and lower edge consist of $2N$ ``$\downarrow$” arrows.  

\noindent (\romannumeral2). The upper edge consists of $2N$ ``$\uparrow$” arrows, and the lower edge consists of $2N$ ``$\downarrow$” arrows. 

\subsubsection{Construction of the dimer lattice}   \label{f-f-a-1}

\begin{figure*} 
\includegraphics{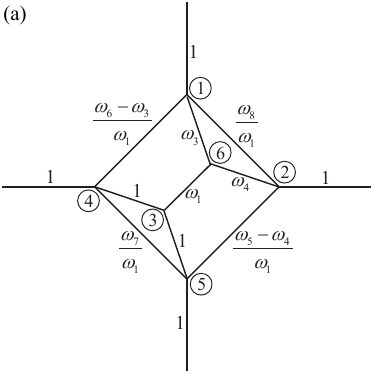} \quad \quad \quad \quad
\includegraphics{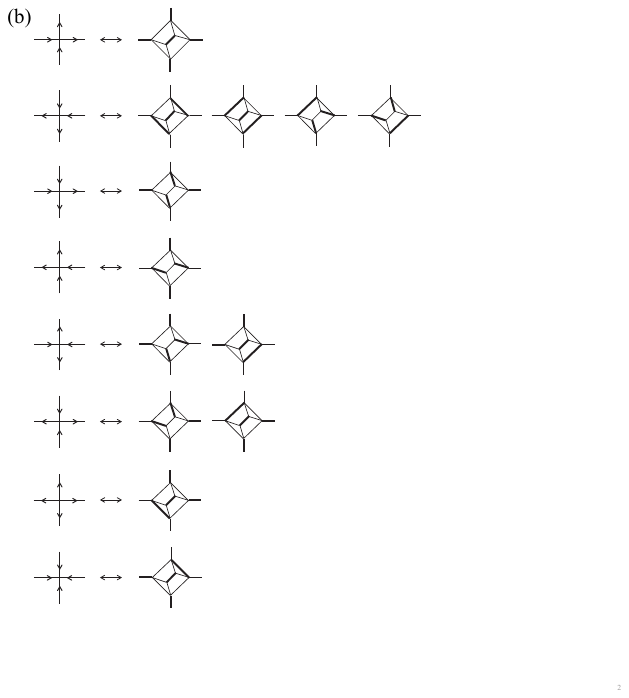}
\caption{The construction of the dimer lattice for the free-fermion model. (a) The dimer cluster for a single vertex site. The associated dimer weights are presented and the six points are numbered. (b) All possible dimer arrangements of a cluster for each of the vertex configurations.} \label{fig2}
\end{figure*}

\begin{figure*} 
\includegraphics{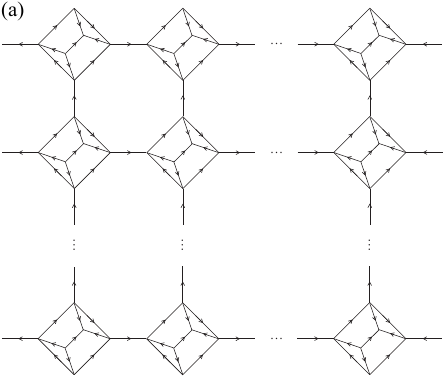} \quad \quad \quad \quad
\includegraphics{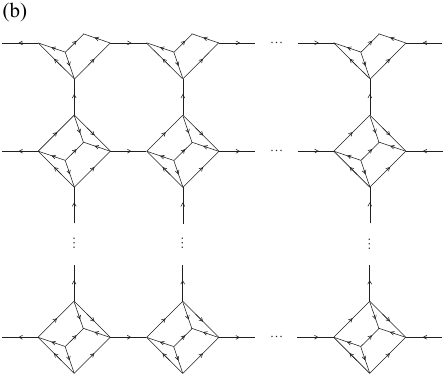}
\caption{The dimer lattices for the free-fermion model under the BCs $(\text{\romannumeral1})$ and $(\text{\romannumeral2})$, with the directions on each edge. (a) The system $(\text{\romannumeral1})^\prime$ corresponding to the BCs $(\text{\romannumeral1})$. (b) The system $(\text{\romannumeral2})^\prime$ corresponding to the BCs $(\text{\romannumeral2})$.} \label{fig3}
\end{figure*}

We adopt the correspondence between the arrow and dimer configurations: ``$\to$'' and ``$\uparrow$'' are regarded as dimers on the horizontal and vertical edges, respectively; correspondingly, ``$\leftarrow$'' and ``$\downarrow$'' indicate the absence of a dimer (i.e., dimer uncovering). Following Ref.~\cite{RN59}, we construct a dimer cluster for each vertex site of the free-fermion model. Figure \ref{fig2}(a) shows the dimer cluster corresponding to a single vertex, along with the assigned dimer weights. Each cluster consists of six points on the dimer lattice, which are numbered in Fig. \ref{fig2}(a). Figure \ref{fig2}(b) lists all possible dimer arrangements for each of the eight-vertex configurations, along with the corresponding arrow-dimer mappings. It can be verified that all vertex weights $\left\{\omega_i,~i = 1, \cdots ,8 \right\}$ are correctly generated. For example, we can examine the weight of vertex (2):
\begin{align*}
&\frac{\omega _8}{\omega _1}{\omega _1}\frac{\omega _7}{\omega _1} + \frac{\omega _6 - \omega _3}{\omega _1}{\omega _1}\frac{\omega _5 - \omega _4}{\omega _1} + \frac{\omega _6 - \omega _3}{\omega _1}{\omega _4} + {\omega _3}\frac{\omega _5 - \omega _4}{\omega _1}   \\
&= \frac{\omega _5\omega _6 + \omega _7\omega _8 - \omega _3\omega _4}{\omega _1}  \\
&= {\omega _2}~.   
\end{align*}
In this way, the partition function of the free-fermion model is translated into that of the corresponding dimer model. The associated BCs, denoted as $(\text{\romannumeral1})^\prime$ and $(\text{\romannumeral2})^\prime$, likewise impose periodic boundary conditions in the $N$ direction and fixed upper and lower edges. These BCs can be readily verified as shown in Fig.~\ref{fig3}. We adopt the convention that the first row is at the top and the first column is on the left.

For the dimer model under $(\text{\romannumeral1})^\prime$ and $(\text{\romannumeral2})^\prime$, either partition function can be expressed as a Pfaffian of a certain matrix. To determine the elements of the matrix, we should first draw directions on each edge of the dimer lattice, to ensure that every closed polygon containing an even number of edges and enclosing an even number of points is clockwise odd \cite{RN397, RN59, RN465}. That is, in a complete circulation around this closed polygon there are an odd number of clockwise directions (and also an odd number of counterclockwise directions). Since both systems $(\text{\romannumeral1})^\prime$ and $(\text{\romannumeral2})^\prime$ are on a cylindrical face and no crossing edges appear, we can make a unique choice of the directions, as shown in Fig. \ref{fig3}. For a detailed analysis of the choice of directions, one can see Sec. IV. 4 of Ref. \cite{RN465}. 

Then we can explicitly define the matrices ${\bf{M}}_1$ for system $(\text{\romannumeral1})^\prime$ and ${\bf{M}}_2$ for system $(\text{\romannumeral2})^\prime$. We number the clusters of the dimer lattice row by row, and the clusters of each row from left to right. Note that there are six sets of lattice sites [see Fig. \ref{fig2}(a)], the sites of each set are numbered in the same way. We use the notation $\left[ i \right]\left(i=1,\cdots,6\right)$ for the sites of each set, for example,
\begin{eqnarray}
\left[ 1 \right] =&&\underbrace{ \textcircled{1}_1 \cdots \textcircled{1}_{2N} }_{1\rm{st}~\rm{row}} \underbrace{ \textcircled{1}_{2N+1} \cdots \textcircled{1}_{4N} }_{2\rm{nd}~\rm{row}} \nonumber \\
&& \cdots \underbrace{ \textcircled{1}_{\left(M-1\right) \times 2N + 1} \cdots \textcircled{1}_{2MN} }_{M\rm{th}~\rm{row}}~.  \label{eq4}
\end{eqnarray}
For system $(\text{\romannumeral1})^\prime$, the matrix ${\bf{M}}_1$ is of the size $12MN\times12MN$ and in a form of
\begin{eqnarray}
{\bf{M}}_1 = \begin{array}{*{20}{l}}
{}~~~~~~~{\begin{array}{*{20}{c}} \left[ 1 \right] \left[ 2 \right] \left[ 3 \right] \left[ 4 \right] \left[ 5 \right] \left[ 6 \right] \end{array}} \\
{\begin{array}{*{20}{c}}
\left[ 1 \right] \\
\left[ 2 \right] \\
\left[ 3 \right] \\
\left[ 4 \right] \\
\left[ 5 \right] \\
\left[ 6 \right]  \end{array}}
{\left[ \begin{array}{*{20}{c}}
{~~~}&{~~}&{~~}&{~~}&{~~}&{~}\\
{}&{}&{}&{}&{}&{}\\
{}&{}&{}&{}&{}&{}\\
{}&{}&{}&{}&{}&{}\\
{}&{}&{}&{}&{}&{}\\
{}&{}&{}&{}&{}&{}
\end{array} \right]}
\end{array}.   \label{eq5}
\end{eqnarray}
Each row of ${\bf{M}}_1$, as well as each column, is labelled with a lattice site. Given two sites $a$ and $b$, the element ${\bf{M}}_1 \left(a,b\right)$ is
\begin{equation}
{\bf{M}}_1\left(a,b\right) = \left\{ \begin{array}{*{20}{l}}
{s\left(a,b\right) \omega \left(a,b\right),~{\rm{if}}~a~{\rm{and}}~b~\rm{are~connected}}\\~~~~~~~~~~~~~~~~~~~~~ {\rm{by~an~edge}} \\
{0,~\rm{otherwise}}
\end{array} \right. ,    \label{eq6}
\end{equation}
where $\omega \left(a,b\right)$ is the dimer weight on the edge connecting $a$ and $b$, and
\begin{equation}
s\left(a,b\right) = \left\{ \begin{array}{*{20}{l}}
{1,~{\rm{if~the~direction~on~the~edge~is~from}}~a~{\rm{to}}~b} \\
{-1,~\rm{otherwise}}
\end{array} \right. .    \label{eq7}
\end{equation}
For example, ${\bf{M}}_1 \left(\textcircled{1}_1, \textcircled{6}_1\right) = \omega_3$, ${\bf{M}}_1 \left(\textcircled{6}_1, \textcircled{1}_1\right) = -\omega_3$ and ${\bf{M}}_1 \left(\textcircled{1}_1, \textcircled{1}_2\right) = 0$. Comparing Fig. \ref{fig3}(b) with Fig. \ref{fig3}(a) we can verify that system $(\text{\romannumeral2})^\prime$ can be obtained by removing $\textcircled{1}_1, \cdots, \textcircled{1}_{2N}$ and the associated edges from system $(\text{\romannumeral1})^\prime$. Therefore, ${\bf{M}}_2$ is actually the part of ${\bf{M}}_1$ with the first $2N$ rows and $2N$ columns removed. ${\bf{M}}_2$ is of the size $(6M-1)2N\times(6M-1)2N$. Both ${\bf{M}}_1$ and ${\bf{M}}_2$ are antisymmetric. The partition function of each system can be evaluated by the Pfaffian of the corresponding matrix:
\begin{align}
&Z_{\left( \text{\romannumeral1} \right)} = Z_{\left( \text{\romannumeral1} \right)^\prime} = {\rm{Pf}}\left( {\bf{M}}_1 \right) = \left[\det \left( {\bf{M}}_1 \right) \right]^{1 \mathord{\left/ {\vphantom {1 2}} \right. \kern-\nulldelimiterspace} 2},  \nonumber \\
&Z_{\left( \text{\romannumeral2} \right)} = Z_{\left( \text{\romannumeral2} \right)^\prime} = {\rm{Pf}}\left( {\bf{M}}_2 \right) = \left[\det \left( {\bf{M}}_2 \right) \right]^{1 \mathord{\left/ {\vphantom {1 2}} \right. \kern-\nulldelimiterspace} 2}.    \label{eq8}
\end{align}

\subsubsection{Pfaffian solution}
We use the Kronecker product to express ${\bf{M}}_1$:
\begin{align}
&{\bf{M}}_1 = {\bf{T}} \otimes {\bf{I}}_M \otimes {\bf{I}}_{2N} + {\bf{A}}_1 \otimes {\bf{I}}_M \otimes {\bf{H}}_{2N} + {\bf{A}}_2 \otimes {\bf{I}}_M \otimes {\bf{H}}_{2N}^T  \nonumber \\
&~~~~~~~~ + {\bf{B}}_1 \otimes {\bf{V}}_M \otimes {\bf{I}}_{2N} + {\bf{B}}_2 \otimes {\bf{V}}_M^T \otimes {\bf{I}}_{2N}~.   \label{eq9}
\end{align}
Here ${\bf{I}}_n$ is the $n \times n$ identity matrix, ${\bf{V}}_M$ and ${\bf{H}}_{2N}$ are $M \times M$ and $2N \times 2N$ matrices, respectively
\begin{align}
{\bf{V}}_M = \left[ \begin{array}{*{20}{c}}
0&0& \cdots &0&0\\
1&0& \cdots &0&0\\
0&1& \cdots &0& 0 \\
 \vdots & \vdots & \ddots & \vdots & \vdots \\
0&0& \cdots &1&0
\end{array} \right],~
{\bf{H}}_{2N} = \left[ \begin{array}{*{20}{c}}
0&1&0& \cdots &0\\
0&0&1& \cdots &0\\
 \vdots & \vdots & \vdots & \ddots &0\\
0&0&0& \cdots &1\\
{ - 1}&0&0& \cdots &0
\end{array} \right],   \label{eq10}
\end{align}
and ${\bf{T}}$, ${\bf{A}}_1$, ${\bf{A}}_2$, ${\bf{B}}_1$ and ${\bf{B}}_2$ are $6\times6$ matrices
\begin{widetext}
\begin{align}
&{\bf{T}} = \left[ {\begin{array}{*{20}{c}}
0&{\frac{{{\omega _8}}}{{{\omega _1}}}}&0&{\frac{{{\omega _3} - {\omega _6}}}{{{\omega _1}}}}&0&{{\omega _3}}\\
{ - \frac{{{\omega _8}}}{{{\omega _1}}}}&0&0&0&{\frac{{{\omega _4} - {\omega _5}}}{{{\omega _1}}}}&{{\omega _4}}\\
0&0&0&1&1&{{\omega _1}}\\
{\frac{{{\omega _6} - {\omega _3}}}{{{\omega _1}}}}&0&{ - 1}&0&{ - \frac{{{\omega _7}}}{{{\omega _1}}}}&0\\
0&{\frac{{{\omega _5} - {\omega _4}}}{{{\omega _1}}}}&{ - 1}&{\frac{{{\omega _7}}}{{{\omega _1}}}}&0&0\\
{ - {\omega _3}}&{ - {\omega _4}}&{ - {\omega _1}}&0&0&0
\end{array}} \right],~
{\bf{A}}_1 = \left[ {\begin{array}{*{20}{c}}
0&0&0&0&0&0\\
0&0&0&1&0&0\\
0&0&0&0&0&0\\
0&0&0&0&0&0\\
0&0&0&0&0&0\\
0&0&0&0&0&0
\end{array}} \right],~
{\bf{B}}_1 = \left[ {\begin{array}{*{20}{c}}
0&0&0&0&1&0\\
0&0&0&0&0&0\\
0&0&0&0&0&0\\
0&0&0&0&0&0\\
0&0&0&0&0&0\\
0&0&0&0&0&0
\end{array}} \right],  \nonumber \\
&{\bf{A}}_2 = -{\bf{A}}_1^T~,~{\bf{B}}_2 = -{\bf{B}}_1^T~.   \label{eq11}
\end{align}
\end{widetext}
The determinant of ${\bf{M}}_1$ can be calculated by block diagonalization. We first observe that ${\bf{H}}_{2N}^T = {\bf{H}}_{2N}^{ - 1}$, thus ${\bf{H}}_{2N}^T$ and ${\bf{H}}_{2N}$ can be simultaneously diagonalized. Explicitly, 
\begin{align}
&~~~~~~{\bf{H}}_{2N} = {\bf{T}}_{\bf{H}}\left[ \begin{array}{*{20}{c}}
{e^{i\alpha _1}}&{}&{}\\
{}& \ddots &{}\\
{}&{}&{e^{i\alpha _{2N}}}
\end{array} \right]{\bf{T}}_{\bf{H}}^{ - 1},  \nonumber  \\
&{\bf{H}}_{2N}^T = {\bf{H}}_{2N}^{ - 1} = {\bf{T}}_{\bf{H}}\left[ \begin{array}{*{20}{c}}
{e^{ - i\alpha _1}}&{}&{}\\
{}& \ddots &{}\\
{}&{}&{e^{ - i\alpha _{2N}}}
\end{array} \right]{\bf{T}}_{\bf{H}}^{ - 1},  \label{eq12}
\end{align}
with $\alpha _i = \frac{\left( 2i - 1 \right)\pi }{2N},~i = 1, \cdots ,2N$. Then we define the matrix $\tilde{\bf{M}}_1$
\begin{align}
\tilde{\bf{M}}_1 & = \left( {\bf{I}}_6 \otimes {\bf{I}}_M \otimes {\bf{T}}_{\bf{H}}^{ - 1} \right) \times {\bf{M}}_1 \times \left( {\bf{I}}_6 \otimes {\bf{I}}_M \otimes {\bf{T}}_{\bf{H}} \right)  \nonumber \\
& = {\bf{T}} \otimes {\bf{I}}_M \otimes {\bf{I}}_{2N} + {\bf{A}}_1 \otimes {\bf{I}}_M \otimes {\rm{diag}}\left\{ e^{i \alpha _i} \right\} + {\bf{A}}_2 \otimes {\bf{I}}_M  \nonumber \\
& \otimes {\rm{diag}}\left\{ e^{-i \alpha _i} \right\} + {\bf{B}}_1 \otimes {\bf{V}}_M \otimes {\bf{I}}_{2N} + {\bf{B}}_2 \otimes {\bf{V}}_M^T \otimes {\bf{I}}_{2N}~,  \label{eq13}
\end{align}
whose determinant equals to that of ${\bf{M}}_1$. By rearranging the rows and columns of $\tilde{\bf{M}}_1$, we obtain a block-diagonal matrix ${\tilde{\tilde{\bf{M}}}}_1$  
\begin{equation}
{\tilde{\tilde{\bf{M}}}}_1 = \left[ \begin{array}{*{20}{c}}
{{\bf{C}}_M \left( \alpha _1 \right)}&{}&{}\\
{}& \ddots &{}\\
{}&{}&{{\bf{C}}_M \left( \alpha _{2N} \right)}
\end{array} \right].   \label{eq14}
\end{equation}   
${\tilde{\tilde{\bf{M}}}}_1$ is a similar matrix of $\tilde{\bf{M}}_1$, thus the determinants of them are identical. In Eq. (\ref{eq14}) the $6M \times 6M$ matrix ${\bf{C}}_M \left( \alpha \right)$ is
\begin{align}
{\bf{C}}_M \left( \alpha \right)& = {\bf{I}}_M \otimes {\bf{T}} + {\bf{I}}_M \otimes \left( e^{i\alpha} {\bf{A}}_1 \right) + {\bf{I}}_M \otimes \left( e^{-i\alpha} {\bf{A}}_2 \right)  \nonumber \\
&~~~~ + {\bf{V}}_M \otimes {\bf{B}}_1 + {\bf{V}}_M^T \otimes {\bf{B}}_2   \nonumber \\
& = \left[ \begin{array}{*{20}{c}}
{\bf{C}}&{{{\bf{B}}_2}}&{\bf{0}}& \cdots &{\bf{0}}&{\bf{0}}&{\bf{0}}\\
{{{\bf{B}}_1}}&{\bf{C}}&{{{\bf{B}}_2}}& \cdots &{\bf{0}}&{\bf{0}}&{\bf{0}}\\
{\bf{0}}&{{{\bf{B}}_1}}&{\bf{C}}& \cdots &{\bf{0}}&{\bf{0}}&{\bf{0}}\\
 \vdots & \vdots & \vdots & \ddots & \vdots & \vdots & \vdots \\
{\bf{0}}&{\bf{0}}&{\bf{0}}& \cdots &{\bf{C}}&{{{\bf{B}}_2}}&{\bf{0}}\\
{\bf{0}}&{\bf{0}}&{\bf{0}}& \cdots &{{{\bf{B}}_1}}&{\bf{C}}&{{{\bf{B}}_2}}\\
{\bf{0}}&{\bf{0}}&{\bf{0}}& \cdots &{\bf{0}}&{{{\bf{B}}_1}}&{\bf{C}}
\end{array} \right],  \label{eq15}
\end{align}
with
\begin{align}
{\bf{C}}& = {\bf{T}} + e^{i\alpha} {\bf{A}}_1 + e^{-i\alpha} {\bf{A}}_2  \nonumber \\
& = \left[ \begin{array}{*{20}{c}}
0&{\frac{{{\omega _8}}}{{{\omega _1}}}}&0&{\frac{{{\omega _3} - {\omega _6}}}{{{\omega _1}}}}&0&{{\omega _3}}\\
{ - \frac{{{\omega _8}}}{{{\omega _1}}}}&0&0&{{e^{i\alpha }}}&{\frac{{{\omega _4} - {\omega _5}}}{{{\omega _1}}}}&{{\omega _4}}\\
0&0&0&1&1&{{\omega _1}}\\
{\frac{{{\omega _6} - {\omega _3}}}{{{\omega _1}}}}&{ - {e^{ - i\alpha }}}&{ - 1}&0&{ - \frac{{{\omega _7}}}{{{\omega _1}}}}&0\\
0&{\frac{{{\omega _5} - {\omega _4}}}{{{\omega _1}}}}&{ - 1}&{\frac{{{\omega _7}}}{{{\omega _1}}}}&0&0\\
{ - {\omega _3}}&{ - {\omega _4}}&{ - {\omega _1}}&0&0&0
\end{array} \right].   \label{eq16}
\end{align}
Applying similar techniques to ${\bf{M}}_2$ leads to
\begin{equation}
\tilde{\bf{M}}_2 = \left( {\bf{I}}_{6M - 1} \otimes {\bf{T}}_{\bf{H}}^{ - 1} \right) \times {\bf{M}}_2 \times \left( {\bf{I}}_{6M - 1} \otimes {\bf{T}}_{\bf{H}} \right)   \label{eq17}
\end{equation}
and 
\begin{equation}
{\tilde{\tilde{\bf{M}}}}_2 = \left[ \begin{array}{*{20}{c}}
{{\bf{C}}_{M,2} \left( \alpha _1 \right)}&{}&{}\\
{}& \ddots &{}\\
{}&{}&{{\bf{C}}_{M,2} \left( \alpha _{2N} \right)}
\end{array} \right].   \label{eq18}
\end{equation}
The $\left(6M-1\right) \times \left(6M-1\right)$ matrix ${\bf{C}}_{M,2} \left( \alpha \right)$ is 
\begin{align}
{\bf{C}}_{M,2} \left( \alpha \right) = \left[ \begin{array}{*{20}{c}}
{\bf{C}}_2&{{{\bf{D}}_2}}&{\bf{0}}& \cdots &{\bf{0}}&{\bf{0}}&{\bf{0}}\\
{{{\bf{D}}_1}}&{\bf{C}}&{{{\bf{B}}_2}}& \cdots &{\bf{0}}&{\bf{0}}&{\bf{0}}\\
{\bf{0}}&{{{\bf{B}}_1}}&{\bf{C}}& \cdots &{\bf{0}}&{\bf{0}}&{\bf{0}}\\
 \vdots & \vdots & \vdots & \ddots & \vdots & \vdots & \vdots \\
{\bf{0}}&{\bf{0}}&{\bf{0}}& \cdots &{\bf{C}}&{{{\bf{B}}_2}}&{\bf{0}}\\
{\bf{0}}&{\bf{0}}&{\bf{0}}& \cdots &{{{\bf{B}}_1}}&{\bf{C}}&{{{\bf{B}}_2}}\\
{\bf{0}}&{\bf{0}}&{\bf{0}}& \cdots &{\bf{0}}&{{{\bf{B}}_1}}&{\bf{C}}
\end{array} \right],  \label{eq19}
\end{align}
with
\begin{align}
&{\bf{C}}_2 = \left[ \begin{array}{*{20}{c}}
0&0&{{e^{i\alpha }}}&{\frac{{{\omega _4} - {\omega _5}}}{{{\omega _1}}}}&{{\omega _4}}\\
0&0&1&1&{{\omega _1}}\\
{ - {e^{ - i\alpha }}}&{ - 1}&0&{ - \frac{{{\omega _7}}}{{{\omega _1}}}}&0\\
{\frac{{{\omega _5} - {\omega _4}}}{{{\omega _1}}}}&{ - 1}&{\frac{{{\omega _7}}}{{{\omega _1}}}}&0&0\\
{ - {\omega _4}}&{ - {\omega _1}}&0&0&0
\end{array} \right],  \nonumber \\
&{\bf{D}}_1 = \left[ \begin{array}{*{20}{c}}
0&0&0&1&0\\
0&0&0&0&0\\
0&0&0&0&0\\
0&0&0&0&0\\
0&0&0&0&0\\
0&0&0&0&0
\end{array} \right],~ {\bf{D}}_2 = -{\bf{D}}_1^T~.  \label{eq20}
\end{align}
That is, ${\bf{C}}_{M,2}\left( \alpha \right)$ is the part of ${\bf{C}}_M \left( \alpha \right)$ with the first row and first column removed. Then we have
\begin{align}
&\det \left( {\bf{M}}_1 \right) = \prod\limits_{i = 1}^{2N} \det \left[ {\bf{C}}_M \left( \alpha _i \right) \right],  \nonumber \\
&\det \left( {\bf{M}}_2 \right) = \prod\limits_{i = 1}^{2N} \det \left[ {\bf{C}}_{M,2} \left( \alpha _i \right) \right].  \label{eq21}
\end{align}
For an exact calculation of $\det \left( {\bf{M}}_1 \right)$ and $\det \left( {\bf{M}}_2 \right)$, we only need to solve two Toeplitz determinants \cite{RN459, RN490} $\det \left[ {\bf{C}}_M \left( \alpha \right) \right]$ and $\det \left[ {\bf{C}}_{M,2} \left( \alpha \right) \right]$. 

We denote $\det \left[ {\bf{C}}_M \left( \alpha \right) \right]$ and $\det \left[ {\bf{C}}_{M,2} \left( \alpha \right) \right]$ by $\left[ {\bf{C}}_M \right]$ and $\left[ {\bf{C}}_{M,2} \right]$, respectively. Regarding the elements 1 in ${\bf{B}}_1$ and $-1$ in ${\bf{B}}_2$, it is clear to verify that $\left[ {\bf{C}}_M \right]$ can be expanded as \cite{RN438, RN382} 
\begin{align*}
\left[ {\bf{C}}_M \right] = \left[ {\bf{C}} \right] \times \left[ {\bf{C}}_{M-1} \right] + \left[ {\bf{C}}_M \right]_{5,7;5,7}~,
\end{align*}
where $\left[ {\bf{C}} \right]$ is $\det \left( {\bf{C}} \right)$ and $\left[ {\bf{C}}_M \right]_{i,j;i,j}$ denotes the determinant of ${\bf{C}}_M \left( \alpha \right)$ with the $i$th and $j$th rows and $i$th and $j$th columns removed. Noticing that $\left[ {\bf{C}}_M \right]_{5,7;5,7} = \left[ {\bf{C}} \right]_{5;5} \times \left[ {\bf{C}}_{M-1,2} \right]$, $\left[ {\bf{C}}_M \right]$ is expressed as 
\begin{equation}
\left[ {\bf{C}}_M \right] = \left[ {\bf{C}} \right] \times \left[ {\bf{C}}_{M-1} \right] + \left[ {\bf{C}} \right]_{5;5} \times \left[ {\bf{C}}_{M-1,2} \right]~.   \label{eq22}
\end{equation}
Similarly, $\left[ {\bf{C}}_{M,2} \right]$ is expanded as
\begin{equation}
\left[ {\bf{C}}_{M,2} \right] = \left[ {\bf{C}} \right]_{1;1} \times \left[ {\bf{C}}_{M-1} \right] + \left[ {\bf{C}} \right]_{1,5;1,5} \times \left[ {\bf{C}}_{M-1,2} \right]~,   \label{eq23}
\end{equation}
where we have used $\left[ {\bf{C}}_2 \right] = \left[ {\bf{C}} \right]_{1;1}$ and $\left[ {\bf{C}}_2 \right]_{4;4} = \left[ {\bf{C}} \right]_{1,5;1,5}$. Notice the initial condition $\left[ {\bf{C}}_1 \right] = \left[ {\bf{C}} \right]$, $\left[ {\bf{C}}_{1,2} \right] = \left[ {\bf{C}} \right]_{1;1}$, and the recursion relation is thus obtained
\begin{align}
\left[ \begin{array}{*{20}{c}}
{\left[ {{{\bf{C}}_M}} \right]}\\
{\left[ {{{\bf{C}}_{M,2}}} \right]}
\end{array} \right] & = \left[ \begin{array}{*{20}{c}}
{\left[ {\bf{C}} \right]}&{{{\left[ {\bf{C}} \right]}_{5;5}}}\\
{{{\left[ {\bf{C}} \right]}_{1;1}}}&{{{\left[ {\bf{C}} \right]}_{1,5;1,5}}}
\end{array} \right] \left[ \begin{array}{*{20}{c}}
{\left[ {{{\bf{C}}_{M - 1}}} \right]}\\
{\left[ {{{\bf{C}}_{M - 1,2}}} \right]}
\end{array} \right]   \nonumber \\
&=  \cdots  \nonumber \\
&= \left[ \begin{array}{*{20}{c}}
{\left[ {\bf{C}} \right]}&{{{\left[ {\bf{C}} \right]}_{5;5}}}\\
{{{\left[ {\bf{C}} \right]}_{1;1}}}&{{{\left[ {\bf{C}} \right]}_{1,5;1,5}}}
\end{array} \right]^{M - 1} \left[ \begin{array}{*{20}{c}}
{\left[ {{{\bf{C}}_1}} \right]}\\
{\left[ {{{\bf{C}}_{1,2}}} \right]}
\end{array} \right]   \nonumber \\
&= \left[ \begin{array}{*{20}{c}}
{\left[ {\bf{C}} \right]}&{{{\left[ {\bf{C}} \right]}_{5;5}}}\\
{{{\left[ {\bf{C}} \right]}_{1;1}}}&{{{\left[ {\bf{C}} \right]}_{1,5;1,5}}}
\end{array} \right]^M \left[ \begin{array}{*{20}{c}}
1\\
0
\end{array} \right]~.  \label{eq24}
\end{align}
Calculate the elements of the recursive matrix and use the notations
\begin{align}
&a = \left[ {\bf{C}} \right] = \omega _2^2 + \omega _3^2 + 2{\omega _2}{\omega _3}\cos \alpha,  \nonumber \\ 
&b = \left[ {\bf{C}} \right]_{5;5} = 2i{\omega _6}{\omega _8}\sin \alpha,  \nonumber \\ 
&c = \left[ {\bf{C}} \right]_{1;1} =  - 2i{\omega _5}{\omega _7}\sin \alpha,  \nonumber \\ 
&d = \left[ {\bf{C}} \right]_{1,5;1,5} = \omega _1^2 + \omega _4^2 - 2{\omega _1}{\omega _4}\cos \alpha.  \label{eq25}
\end{align}
The eigenvalues of the recursive matrix are
\begin{equation}
\lambda _{1,2} = \frac{1}{2}\left[ a + d \pm \sqrt {{\left( a + d \right)}^2 - 4\left( ad - bc \right)}  \right].  \label{eq26}
\end{equation}

To solve the recursion relation, one can use the method introduced in Refs. \cite{RN438} and \cite{RN382}, which employs the generating functions of $\left[ {\bf{C}}_M \right]$ and $\left[ {\bf{C}}_{M,2} \right]$. Or we can directly calculate the recursive matrix to the $M$th power by using
\begin{align}
\left[ \begin{array}{*{20}{c}} a&b \\ c&d \end{array} \right] \left[ \begin{array}{*{20}{c}} b&b \\ {\lambda _1 - a}&{\lambda _2 - a} \end{array} \right]
 = \left[ \begin{array}{*{20}{c}} b&b \\ {\lambda _1 - a}&{\lambda _2 - a} \end{array} \right] \left[ \begin{array}{*{20}{c}} {\lambda _1}&{} \\ {}&{\lambda _2} \end{array} \right].  \label{eq27}
\end{align}
This yields the result 
\begin{subequations}
\label{eq28}
\begin{eqnarray}
\left[ {\bf{C}}_M \right] = \frac{\lambda _1^{M + 1} - \lambda _2^{M + 1}}{\lambda _1 - \lambda _2} - d\frac{\lambda _1^M - \lambda _2^M}{\lambda _1 - \lambda _2}~,  \label{eq28a}
\end{eqnarray}
\begin{eqnarray}
\left[ {\bf{C}}_{M,2} \right] = c\frac{\lambda _1^M - \lambda _2^M}{\lambda _1 - \lambda _2}~.  \label{eq28b}
\end{eqnarray}
\end{subequations}
As shown in Refs. \cite{RN410} and \cite{RN432}, the term $\lambda _1^M - \lambda _2^M$ allows us to write the solution in a product form. To indicate this in detail, we can make use of an identity
\begin{align}
&\gamma ^{M - 1} + \gamma ^{M - 3} + \cdots + \gamma ^{ - \left( M - 3 \right)} + \gamma ^{ - \left( M - 1 \right)}  \nonumber \\
= &\prod\limits_{j = 1}^{M - 1} \left( \gamma + \gamma ^{-1} - 2\cos \frac{j\pi }{M} \right)~.   \label{eq29}
\end{align}
This identity can be easily proved by verifying that both the left-hand and right-hand sides have the same roots $\left\{ e^{ \pm i\frac{j\pi }{M}},~j = 1, \cdots ,M - 1 \right\}$. Equation (\ref{eq29}) leads to 
\begin{align}
\!\!\!\!\!\! \gamma ^M - \gamma ^{-M} = \left( \gamma - \gamma ^{-1} \right)\prod\limits_{j = 1}^{M - 1} \left( \gamma + \gamma ^{-1} - 2\cos \frac{j\pi }{M} \right).  \label{eq30}
\end{align}
Now we are able to express $\left[ {\bf{C}}_{M,2} \right]$ in Eq. (\ref{eq28b}) as
\begin{align}
\left[ {\bf{C}}_{M,2} \right] &= c \frac{{\left( \lambda _1 \lambda _2 \right)}^{M \mathord{\left/ {\vphantom {M 2}} \right. \kern-\nulldelimiterspace} 2}}{\lambda _1 - \lambda _2} \left[ {\left( \frac{\lambda _1}{\lambda _2} \right)}^{M \mathord{\left/ {\vphantom {M 2}} \right. \kern-\nulldelimiterspace} 2} - {\left( \frac{\lambda _2}{\lambda _1} \right)}^{M \mathord{\left/ {\vphantom {M 2}} \right. \kern-\nulldelimiterspace} 2} \right]   \nonumber \\
&= c \frac{{\left( \lambda _1 \lambda _2 \right)}^{M \mathord{\left/ {\vphantom {M 2}} \right. \kern-\nulldelimiterspace} 2}}{\lambda _1 - \lambda _2} \left[ {\left( \frac{\lambda _1}{\lambda _2} \right)}^{1/2} - {\left( \frac{\lambda _2}{\lambda _1} \right)}^{1/2} \right]   \nonumber \\
&~~~\times \prod\limits_{j = 1}^{M - 1} \left[ {\left( \frac{\lambda _1}{\lambda _2} \right)}^{1/2} + {\left( \frac{\lambda _2}{\lambda _1} \right)}^{1/2} - 2\cos \frac{j\pi}{M} \right]   \nonumber \\
& = c\prod\limits_{j = 1}^{M - 1} \left[ \lambda _1 + \lambda _2 - 2{\left( \lambda _1 \lambda _2 \right)}^{1/2} \cos \frac{j\pi}{M} \right]  \nonumber \\
& = c\prod\limits_{j = 1}^{M - 1} \left[ a + d - 2{\left( ad - bc \right)}^{1/2} \cos \frac{j\pi}{M} \right]~.  \label{eq31}
\end{align}
Now we see $\left[ {\bf{C}}_{M,2} \right]$ is our choice as it has a single product form. 

The last step of our derivation is to substitute Eq. (\ref{eq31}) into Eqs. (\ref{eq21}) and (\ref{eq8}). The following calculation 
\begin{align}
\left( -1 \right)^N \prod\limits_{i = 1}^{2N} \sin {\alpha _i}  = \left[ \prod\limits_{i = 1}^N \sin \frac{\left( 2i - 1 \right)\pi }{2N}  \right]^2 = \frac{1}{2^{2\left( N - 1 \right)}}  \label{eq32}
\end{align}
yields
\begin{align}
\prod\limits_{i = 1}^{2N} c\left( \alpha _i \right)  = 2^2 \left( \omega _5 \omega _7 \right)^{2N}~.  \label{eq33}
\end{align}
By using Eq. (\ref{eq33}) and verifying that the term $\left[ \cdots \right]$ in Eq. (\ref{eq31}) has the symmetric property $\left[ \cdots \right]\left( \alpha _i \right) = \left[ \cdots \right]\left( \alpha _{2N - i + 1} \right)$, we achieve the expression of $Z_{\left( \text{\romannumeral2} \right)}$
\begin{align}
\!\!\!Z_{\left( \text{\romannumeral2} \right)} = 2 \left( \omega _5 \omega _7 \right)^N \prod\limits_{i = 1}^N \prod\limits_{j = 1}^{M - 1} \left[ a_i + d_i - 2{\left( a_i d_i - b_i c_i \right)}^{1/2} \cos \frac{j\pi}{M} \right]  \label{eq34} 
\end{align}
with
\begin{align}
\!\!\!\!\!&a_i + d_i = \omega _1^2 + \omega _2^2 + \omega _3^2 + \omega _4^2 + 2\left( \omega _2 \omega _3 - \omega _1 \omega _4 \right) \cos \frac{\left( 2i - 1 \right)\pi}{2N},  \nonumber \\
\!\!\!\!\!&a_i d_i - b_i c_i = \left[ \omega _1 \omega _3 - \omega _2 \omega _4 + \left( \omega _1 \omega _2 - \omega _3 \omega _4 \right)\cos \frac{\left( 2i - 1 \right)\pi}{2N} \right]^2  \nonumber \\
\!\!\!\!\!&~~~~~~~~~~~~~~~~~~ + \left( \omega _5 \omega _6 - \omega _7 \omega _8 \right)^2 \sin ^2 \frac{\left( 2i - 1 \right)\pi}{2N}.   \label{eq35}
\end{align}
This result demonstrates that we have succeeded in finding the BCs $\left( \text{\romannumeral2} \right)$ for the free-fermion model, under which the partition function of finite lattice has a double product form. 

\subsubsection{Thermodynamic limit}
Equation (\ref{eq34}) can be rearranged as
\begin{widetext}
\begin{align}
Z_{\left( \text{\romannumeral2} \right)}& = 2 \left( \omega _5 \omega _7 \right)^N \prod\limits_{i = 1}^N \prod\limits_{j = 1}^{M - 1} {\left[ \left( a_i + d_i \right)^2 - 4\left( a_i d_i - b_i c_i \right) \cos ^2 \frac{j\pi}{M} \right]}^{1/2}   \nonumber \\
&= 2 \left( \omega _5 \omega _7 \right)^N \prod\limits_{i = 1}^N \prod\limits_{j = 1}^{M - 1} \left[ \bar A + \bar B \cos \frac{2j\pi}{M} + \bar C \cos \frac{\left( 2i - 1 \right)\pi}{2N} + \bar D \cos \frac{\left( 2i - 1 \right)\pi}{N} + \bar E \cos \frac{2j\pi}{M} \cos \frac{\left( 2i - 1 \right)\pi}{2N}  \right.   \nonumber \\
&~~~~~~~~~~~~~~~~~~~~~~~~~~~~~~~~~~~~~~~~~~~~~~~~~~~~~~~~~~~~~~~~~~~~~~~~~~~~~~~~~~~~~~~~~~~~~~~~~\left. + \bar F \cos \frac{2j\pi}{M} \cos \frac{\left( 2i - 1 \right)\pi}{N} \right]^{1/2},     \label{eq36} 
\end{align}
with
\begin{align}
&\bar A = \left( \omega _1^2 + \omega _4^2 \right)^2 + \left( \omega _2^2 + \omega _3^2 \right)^2 + 2\left( \omega _1^2 \omega _4^2 + \omega _2^2 \omega _3^2 \right) + 4\omega _5 \omega _6 \omega _7 \omega _8~,   \nonumber \\
&\bar B =  - 2\left( {\omega _1^2 + \omega _4^2} \right)\left( {\omega _2^2 + \omega _3^2} \right) + 4\left( {{\omega _1}{\omega _2}{\omega _3}{\omega _4} + {\omega _5}{\omega _6}{\omega _7}{\omega _8}} \right)~,  \nonumber \\
&\bar C = 4{\omega _2}{\omega _3}\left( {\omega _2^2 + \omega _3^2} \right) - 4{\omega _1}{\omega _4}\left( {\omega _1^2 + \omega _4^2} \right)~,  \nonumber \\
&\bar D = 2\left( {\omega _1^2\omega _4^2 + \omega _2^2\omega _3^2} \right) - 4{\omega _5}{\omega _6}{\omega _7}{\omega _8}~,  \nonumber \\
&\bar E =  - 4\left( {{\omega _1}{\omega _2} - {\omega _3}{\omega _4}} \right)\left( {{\omega _1}{\omega _3} - {\omega _2}{\omega _4}} \right)~,  \nonumber \\
&\bar F = 4\left( {{\omega _1}{\omega _2}{\omega _3}{\omega _4} - {\omega _5}{\omega _6}{\omega _7}{\omega _8}} \right)~.   \label{eq37}
\end{align}
Then the solution in the thermodynamic limit $M,N \to \infty$ is obtained
\begin{align}
\mathop {\lim }\limits_{M,N \to \infty} \frac{1}{2MN} \ln Z_{\left( \text{\romannumeral2} \right)} = \frac{1}{16 \pi ^2} \int_0^{2\pi } d\theta \int_0^{2\pi} d\phi \ln \left[ \bar A + \bar B\cos \theta + \bar C\cos \phi + \bar D\cos 2\phi + \bar E\cos \theta \cos \phi + \bar F\cos \theta \cos 2\phi  \right].   \label{eq38}
\end{align}
\end{widetext}
Different expressions of this solution can be seen in Refs. \cite{RN65} and \cite{RN59}, and Appendices E.3 and E.4 of Ref. \cite{RN124}.

In the end of this section, we outline the main result. For a square lattice of $M$ rows and $2N$ columns, we have found the BCs---periodic boundary conditions in the $N$ direction, $2N$ ``$\uparrow$'' arrows on the upper edge, and $2N$ ``$\downarrow$'' arrows on the lower edge. Under these BCs the partition function of the free-fermion model has an expression of a double product form, as shown in Eq. (\ref{eq34}) or Eq. (\ref{eq36}).

\subsection{Row-by-row staggered case}   \label{f-f-b}
In this section we consider the free-fermion model with row-by-row staggered weights. There are two sublattices $A$ and $B$ in a row-by-row pattern, and two sets of weights $\left\{ \omega _{iA},~i = 1, \cdots ,8 \right\}$ and $\left\{ \omega _{iB},~i = 1, \cdots ,8 \right\}$ are attached to the sites on $A$ and $B$, respectively. Both $\left\{ \omega _{iA},~i = 1, \cdots ,8 \right\}$ and $\left\{ \omega _{iB},~i = 1, \cdots ,8 \right\}$ satisfy the free-fermion condition [Eq. (\ref{eq2})]. The studied system is still a finite square lattice of $M$ rows and $2N$ columns. Below we show that in this case the solution under the BCs $\left( \text{\romannumeral2} \right)$ can also be given in a product form, when $M$ is even. We set the first row at the top to be on sublattice $A$, thus the last row is on sublattice $B$.

\begin{figure} 
\includegraphics{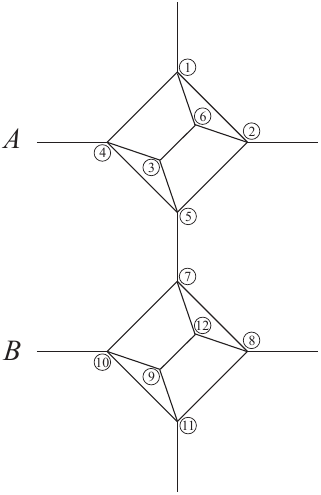}
\caption{The dimer cluster for the row-by-row staggered free-fermion model, with the top vertex site on sublattice $A$ and the bottom one on $B$. Twelve points are numbered.} \label{fig4}
\end{figure}

We still use the Pfaffian method to solve this case. The dimer lattice, as well as the determination of the directions on each edge, is the same as that for the uniform case, see Fig. \ref{fig3}. But in this case each dimer cluster consists of a couple of vertex sites in a column, as shown in Fig. \ref{fig4}. Each cluster can be regarded as a basic unit, forming an $M/2 \times 2N$ square lattice. There are twelve points of the dimer lattice in each cluster, which are numbered in Fig. \ref{fig4}. Now we see that there are also two sets of dimer weights, corresponding to two subclusters $A$ and $B$, respectively. In each subcluster, the dimer weights are determined in the same way as that in the uniform case, see Fig. \ref{fig2}(a). The BCs of the dimer model are still denoted by $(\text{\romannumeral1})^\prime$ and $(\text{\romannumeral2})^\prime$.

Then we define the $12MN\times12MN$ antisymmetric matrix ${\bf{M}}_{{\rm{sta}},1}$ and the $(6M-1)2N\times(6M-1)2N$ antisymmetric matrix ${\bf{M}}_{{\rm{sta}},2}$ in the same way as we have done for ${\bf{M}}_{1}$ and ${\bf{M}}_{2}$, see Eqs. (\ref{eq4})--(\ref{eq7}). The desired solution is again expressed as the Pfaffian of ${\bf{M}}_{{\rm{sta}},2}$
\begin{align}
\!\!\!\!\! Z_{{\rm{sta}},\left( \text{\romannumeral2} \right)} = Z_{{\rm{sta}},\left( \text{\romannumeral2} \right)^\prime} = {\rm{Pf}}\left( {\bf{M}}_{{\rm{sta}},2} \right) = \left[\det \left( {\bf{M}}_{{\rm{sta}},2} \right) \right]^{1/2}.    \label{eq39}
\end{align}

\subsubsection{Pfaffian solution}
Similarly, ${\bf{M}}_{{\rm{sta}},1}$ can be expressed using the Kronecker product
\begin{align}
{\bf{M}}_{{\rm{sta}},1}& = {\bf{T}}_{\rm{sta}} \otimes {\bf{I}}_{M/2} \otimes {\bf{I}}_{2N} + {\bf{A}}_{{\rm{sta}},1} \otimes {\bf{I}}_{M/2} \otimes {\bf{H}}_{2N}   \nonumber \\
&~~~~ + {\bf{A}}_{{\rm{sta}},2} \otimes {\bf{I}}_{M/2} \otimes {\bf{H}}_{2N}^T + {\bf{B}}_{{\rm{sta}},1} \otimes {\bf{V}}_{M/2} \otimes {\bf{I}}_{2N}  \nonumber \\
&~~~~ + {\bf{B}}_{{\rm{sta}},2} \otimes {\bf{V}}_{M/2}^T \otimes {\bf{I}}_{2N}~,   \label{eq40}
\end{align}
where
\begin{widetext}
\begin{align}
{\bf{T}}_{\rm{sta}}& = \left[ \begin{array}{*{20}{c}}
0&{\frac{{{\omega _{8A}}}}{{{\omega _{1A}}}}}&0&{\frac{{{\omega _{3A}} - {\omega _{6A}}}}{{{\omega _{1A}}}}}&0&{{\omega _{3A}}}&0&0&0&0&0&0\\
{ - \frac{{{\omega _8}}}{{{\omega _1}}}}&0&0&0&{\frac{{{\omega _{4A}} - {\omega _{5A}}}}{{{\omega _{1A}}}}}&{{\omega _{4A}}}&0&0&0&0&0&0\\
0&0&0&1&1&{{\omega _{1A}}}&0&0&0&0&0&0\\
{\frac{{{\omega _{6A}} - {\omega _{3A}}}}{{{\omega _{1A}}}}}&0&{ - 1}&0&{ - \frac{{{\omega _{7A}}}}{{{\omega _{1A}}}}}&0&0&0&0&0&0&0\\
0&{\frac{{{\omega _{5A}} - {\omega _{4A}}}}{{{\omega _{1A}}}}}&{ - 1}&{\frac{{{\omega _{7A}}}}{{{\omega _{1A}}}}}&0&0&{ - 1}&0&0&0&0&0\\
{ - {\omega _{3A}}}&{ - {\omega _{4A}}}&{ - {\omega _{1A}}}&0&0&0&0&0&0&0&0&0\\
0&0&0&0&1&0&0&{\frac{{{\omega _{8B}}}}{{{\omega _{1B}}}}}&0&{\frac{{{\omega _{3B}} - {\omega _{6B}}}}{{{\omega _{1B}}}}}&0&{{\omega _{3B}}}\\
0&0&0&0&0&0&{ - \frac{{{\omega _{8B}}}}{{{\omega _{1B}}}}}&0&0&0&{\frac{{{\omega _{4B}} - {\omega _{5B}}}}{{{\omega _{1B}}}}}&{{\omega _{4B}}}\\
0&0&0&0&0&0&0&0&0&1&1&{{\omega _{1B}}}\\
0&0&0&0&0&0&{\frac{{{\omega _{6B}} - {\omega _{3B}}}}{{{\omega _{1B}}}}}&0&{ - 1}&0&{ - \frac{{{\omega _{7B}}}}{{{\omega _{1B}}}}}&0\\
0&0&0&0&0&0&0&{\frac{{{\omega _{5B}} - {\omega _{4B}}}}{{{\omega _{1B}}}}}&{ - 1}&{\frac{{{\omega _{7B}}}}{{{\omega _{1B}}}}}&0&0\\
0&0&0&0&0&0&{ - {\omega _{3B}}}&{ - {\omega _{4B}}}&{ - {\omega _{1B}}}&0&0&0
\end{array} \right]  \nonumber \\
& = \left[ \begin{array}{*{20}{c}}
{{\bf{T}}_A}&{{\bf{B}}_2} \\
{{\bf{B}}_1}&{{\bf{T}}_B}
\end{array} \right],     \label{eq41}
\end{align}
and
\begin{align}
&{\bf{A}}_{{\rm{sta}},1} = \left[ \begin{array}{*{20}{c}}
0&0&0&0&0&0&0&0&0&0&0&0\\
0&0&0&1&0&0&0&0&0&0&0&0\\
0&0&0&0&0&0&0&0&0&0&0&0\\
0&0&0&0&0&0&0&0&0&0&0&0\\
0&0&0&0&0&0&0&0&0&0&0&0\\
0&0&0&0&0&0&0&0&0&0&0&0\\
0&0&0&0&0&0&0&0&0&0&0&0\\
0&0&0&0&0&0&0&0&0&1&0&0\\
0&0&0&0&0&0&0&0&0&0&0&0\\
0&0&0&0&0&0&0&0&0&0&0&0\\
0&0&0&0&0&0&0&0&0&0&0&0\\
0&0&0&0&0&0&0&0&0&0&0&0
\end{array} \right] = \left[ \begin{array}{*{20}{c}}
{{\bf{A}}_1}&{\bf{0}} \\
{\bf{0}}&{{\bf{A}}_1}
\end{array} \right],~ 
{\bf{B}}_{{\rm{sta}},1} = \left[ \begin{array}{*{20}{c}}
0&0&0&0&0&0&0&0&0&0&1&0\\
0&0&0&0&0&0&0&0&0&0&0&0\\
0&0&0&0&0&0&0&0&0&0&0&0\\
0&0&0&0&0&0&0&0&0&0&0&0\\
0&0&0&0&0&0&0&0&0&0&0&0\\
0&0&0&0&0&0&0&0&0&0&0&0\\
0&0&0&0&0&0&0&0&0&0&0&0\\
0&0&0&0&0&0&0&0&0&0&0&0\\
0&0&0&0&0&0&0&0&0&0&0&0\\
0&0&0&0&0&0&0&0&0&0&0&0\\
0&0&0&0&0&0&0&0&0&0&0&0\\
0&0&0&0&0&0&0&0&0&0&0&0
\end{array} \right] = \left[ \begin{array}{*{20}{c}}
{\bf{0}}&{{\bf{B}}_1} \\
{\bf{0}}&{\bf{0}}
\end{array} \right],  \nonumber \\
&{\bf{A}}_{{\rm{sta}},2} = -{\bf{A}}_{{\rm{sta}},1}^T~,~{\bf{B}}_{{\rm{sta}},2} = -{\bf{B}}_{{\rm{sta}},1}^T~.    \label{eq42}
\end{align}
\end{widetext}
${\bf{T}}_A$ and ${\bf{T}}_B$ are defined in Eq. (\ref{eq11}) corresponding to two sets of weights $\left\{ \omega_{iA} \right\}$ and $\left\{ \omega_{iB} \right\}$, respectively. ${\bf{M}}_{{\rm{sta}},2}$ is the part of ${\bf{M}}_{{\rm{sta}},1}$ with the first $2N$ rows and $2N$ columns removed. Employing the techniques similar to those for ${\bf{M}}_{1}$ and ${\bf{M}}_{2}$, we can obtain the block diagonal matrices
\begin{equation}
{\tilde{\tilde{\bf{M}}}}_{{\rm{sta}},1} = \left[ \begin{array}{*{20}{c}}
{{\bf{C}}_{{\rm{sta}},M/2} \left( \alpha _1 \right)}&{}&{}\\
{}& \ddots &{}\\
{}&{}&{{\bf{C}}_{{\rm{sta}},M/2} \left( \alpha _{2N} \right)}
\end{array} \right]   \label{eq43}
\end{equation}
and
\begin{equation}
\!\!\!\!\! {\tilde{\tilde{\bf{M}}}}_{{\rm{sta}},2} = \left[ \begin{array}{*{20}{c}}
{{\bf{C}}_{{\rm{sta}},M/2,2} \left( \alpha _1 \right)}&{}&{}\\
{}& \ddots &{}\\
{}&{}&{{\bf{C}}_{{\rm{sta}},M/2,2} \left( \alpha _{2N} \right)}
\end{array} \right],   \label{eq44}
\end{equation}
which have the same determinants with ${\bf{M}}_{{\rm{sta}},1}$ and ${\bf{M}}_{{\rm{sta}},2}$, respectively. Here
\begin{align}
&{\bf{C}}_{{\rm{sta}},M/2} \left( \alpha \right)   \nonumber \\
&~~~~ = \left[ \begin{array}{*{20}{c}}
{{\bf{C}}_{\rm{sta}}}&{{\bf{B}}_{\rm{sta},2}}&{\bf{0}}& \cdots &{\bf{0}}&{\bf{0}}&{\bf{0}}\\
{{\bf{B}}_{\rm{sta},1}}&{{\bf{C}}_{\rm{sta}}}&{{\bf{B}}_{\rm{sta},2}}& \cdots &{\bf{0}}&{\bf{0}}&{\bf{0}}\\
{\bf{0}}&{{\bf{B}}_{\rm{sta},1}}&{{\bf{C}}_{\rm{sta}}}& \cdots &{\bf{0}}&{\bf{0}}&{\bf{0}}\\
 \vdots & \vdots & \vdots & \ddots & \vdots & \vdots & \vdots \\
{\bf{0}}&{\bf{0}}&{\bf{0}}& \cdots &{{\bf{C}}_{\rm{sta}}}&{{\bf{B}}_{\rm{sta},2}}&{\bf{0}}\\
{\bf{0}}&{\bf{0}}&{\bf{0}}& \cdots &{{\bf{B}}_{\rm{sta},1}}&{{\bf{C}}_{\rm{sta}}}&{{\bf{B}}_{\rm{sta},2}}\\
{\bf{0}}&{\bf{0}}&{\bf{0}}& \cdots &{\bf{0}}&{{\bf{B}}_{\rm{sta},1}}&{{\bf{C}}_{\rm{sta}}}
\end{array} \right]   \label{eq45}
\end{align}
with
\begin{align}
{\bf{C}}_{\rm{sta}} = \left[ {\begin{array}{*{20}{c}}
{{\bf{C}}_A}&{{\bf{B}}_2}\\
{{\bf{B}}_1}&{{\bf{C}}_B}
\end{array}} \right],    \label{eq46}
\end{align}
and ${\bf{C}}_A$ and ${\bf{C}}_B$ are defined in Eq. (\ref{eq16}). ${\bf{C}}_{{\rm{sta}},M/2,2}\left( \alpha \right)$ is the part of ${\bf{C}}_{{\rm{sta}},M/2} \left( \alpha \right)$ with the first row and first column removed. We still have
\begin{align}
\det \left( {\bf{M}}_{{\rm{sta}},2} \right) = \prod\limits_{i = 1}^{2N} \det \left[ {\bf{C}}_{{\rm{sta}},M/2,2} \left( \alpha _i \right) \right].  \label{eq47}
\end{align}

Again we obtain the recursion relation for the Toeplitz determinants $\left[ {\bf{C}}_{{\rm{sta}},M/2} \right]$ and $\left[ {\bf{C}}_{{\rm{sta}},M/2,2} \right]$, similar to Eq. (\ref{eq24}):
\begin{align}
\left[ \begin{array}{*{20}{c}}
{\left[ {\bf{C}}_{{\rm{sta}},M/2} \right]}\\
{\left[ {{{\bf{C}}_{{\rm{sta}},M/2,2}}} \right]}
\end{array} \right]& = \left[ \begin{array}{*{20}{c}}
{\left[ {{{\bf{C}}_{{\rm{sta}}}}} \right]}&{{{\left[ {{{\bf{C}}_{{\rm{sta}}}}} \right]}_{11;11}}}\\
{{{\left[ {{{\bf{C}}_{{\rm{sta}}}}} \right]}_{1;1}}}&{{{\left[ {{{\bf{C}}_{{\rm{sta}}}}} \right]}_{1,11;1,11}}}
\end{array} \right]^{M/2} \left[ \begin{array}{*{20}{c}}
1\\
0
\end{array} \right]   \nonumber \\
& \equiv \left[ \begin{array}{*{20}{c}}
{a_{\rm{sta}}}&{b_{\rm{sta}}}\\
{c_{\rm{sta}}}&{d_{\rm{sta}}}
\end{array} \right]^{M/2} \left[ \begin{array}{*{20}{c}}
1\\
0
\end{array} \right].  \label{eq48}
\end{align}
In the last step we use the notations similar to Eq. (\ref{eq25}), for the elements of the recursive matrix. From Eq. (\ref{eq31}) we immediately know that
\begin{align}
&\left[ {\bf{C}}_{{\rm{sta}},M/2,2} \right] = c_{\rm{sta}} \prod\limits_{j = 1}^{M/2 - 1} \left[ a_{\rm{sta}} + d_{\rm{sta}}  \right.   \nonumber \\
&~~~~~~~~~~~~~~~~ \left. - 2{\left( a_{\rm{sta}} d_{\rm{sta}} - b_{\rm{sta}} c_{\rm{sta}} \right)}^{1/2} \cos \frac{2j\pi}{M} \right]~.  \label{eq49}
\end{align}
It is straightforward to examine the relation between $\left\{ a_{\rm{sta}}, b_{\rm{sta}}, c_{\rm{sta}}, d_{\rm{sta}} \right\}$ and $\left\{ a, b, c, d \right\}$
\begin{align}
\left[ \begin{array}{*{20}{c}}
{{a_{{\rm{sta}}}}}&{{b_{{\rm{sta}}}}}\\
{{c_{{\rm{sta}}}}}&{{d_{{\rm{sta}}}}}
\end{array} \right] = \left[ \begin{array}{*{20}{c}}
{{a_A}}&{{b_A}}\\
{{c_A}}&{{d_A}}
\end{array} \right]\left[ \begin{array}{*{20}{c}}
{{a_B}}&{{b_B}}\\
{{c_B}}&{{d_B}}
\end{array} \right]  \label{eq50}
\end{align}
from Eq. (\ref{eq46}). Then we see that when $\left\{ \omega_A \right\}=\left\{ \omega_B \right\}$, i.e., the system is reduced to the uniform case, Eqs. (\ref{eq48}) and (\ref{eq49}) become Eqs. (\ref{eq24}) and (\ref{eq31}), respectively.

Now we calculate $\left\{ a_{\rm{sta}}, b_{\rm{sta}}, c_{\rm{sta}}, d_{\rm{sta}} \right\}$ explicitly
\begin{widetext}
\begin{align}
&a_{\rm{sta}} = \left( \omega _{2A}^2 + \omega _{3A}^2 \right)\left( \omega _{2B}^2 + \omega _{3B}^2 \right) + 2\left( \omega _{2A} \omega _{3A} \omega _{2B} \omega _{3B} + \omega _{6A} \omega _{8A} \omega _{5B} \omega _{7B} \right) + 2\left( \omega _{2A} \omega _{3B} + \omega _{3A} \omega _{2B} \right)    \nonumber \\
&~~~~~~~~~\times \left( \omega _{2A} \omega _{2B} + \omega _{3A} \omega _{3B} \right)\cos \alpha + 2\left( \omega _{2A} \omega _{3A} \omega _{2B} \omega _{3B} - \omega _{6A} \omega _{8A} \omega _{5B} \omega _{7B} \right)\cos 2\alpha~,  \nonumber \\
&b_{\rm{sta}} = 2i\sin \alpha \left[ \left( \omega _{2A}^2 + \omega _{3A}^2 \right) \omega _{6B} \omega _{8B} + \left( \omega _{1B}^2 + \omega _{4B}^2 \right) \omega _{6A} \omega _{8A} - 2\left( \omega _{6A} \omega _{8A} \omega _{1B} \omega _{4B} - \omega _{2A} \omega _{3A} \omega _{6B} \omega _{8B} \right)\cos \alpha  \right],   \nonumber \\
&c_{\rm{sta}} = -2i\sin \alpha \left[ \left( \omega _{1A}^2 + \omega _{4A}^2 \right) \omega _{5B} \omega _{7B} + \left( \omega _{2B}^2 + \omega _{3B}^2 \right) \omega _{5A} \omega _{7A} - 2\left( \omega _{1A} \omega _{4A} \omega _{5B} \omega _{7B} - \omega _{5A} \omega _{7A} \omega _{2B} \omega _{3B} \right)\cos \alpha  \right],   \nonumber \\
&d_{\rm{sta}} = \left( \omega _{1A}^2 + \omega _{4A}^2 \right)\left( \omega _{1B}^2 + \omega _{4B}^2 \right) + 2\left( \omega _{1A} \omega _{4A} \omega _{1B} \omega _{4B} + \omega _{5A} \omega _{7A} \omega _{6B} \omega _{8B} \right) - 2\left( \omega _{1A} \omega _{4B} + \omega _{4A} \omega _{1B} \right)    \nonumber \\
&~~~~~~~~~\times \left( \omega _{1A} \omega _{1B} + \omega _{4A} \omega _{4B} \right)\cos \alpha + 2\left( \omega _{1A} \omega _{4A} \omega _{1B} \omega _{4B} - \omega _{5A} \omega _{7A} \omega _{6B} \omega _{8B} \right)\cos 2\alpha~.   \label{eq51}
\end{align}
Note that the term $\left[ \cdots \right]$ in Eq. (\ref{eq49}) has the symmetric property $\left[ \cdots \right]\left( \alpha _i \right) = \left[ \cdots \right]\left( \alpha _{2N - i + 1} \right)$. Substituting Eq. (\ref{eq49}) into Eqs. (\ref{eq47}) and (\ref{eq39}), and using Eq. (\ref{eq32}) and $a_{\rm{sta}} d_{\rm{sta}} - b_{\rm{sta}} c_{\rm{sta}} = \left( ad-bc \right)_A \left( ad-bc \right)_B$, we present the expression of $Z_{{\rm{sta}},\left( \text{\romannumeral2} \right)}$
\begin{align}
&Z_{{\rm{sta}},\left( \text{\romannumeral2} \right)} = 2\prod\limits_{i = 1}^{N} \left\{ \left[ \left( \omega _{1A}^2 + \omega _{4A}^2 \right) \omega _{5B} \omega _{7B} + \left( \omega _{2B}^2 + \omega _{3B}^2 \right) \omega _{5A} \omega _{7A} - 2\left( \omega _{1A} \omega _{4A} \omega _{5B} \omega _{7B} - \omega _{5A} \omega _{7A} \omega _{2B} \omega _{3B} \right) \cos {\alpha _i} \right] \right.   \nonumber \\
&~~~~~~~~~~~~~~~~~~~~~\left. \times \prod\limits_{j = 1}^{M/2 - 1} \left[ a_{{\rm{sta}},i} + d_{{\rm{sta}},i} - 2\left( a_i d_i - b_i c_i \right)_A^{1/2} \left( a_i d_i - b_i c_i \right)_B^{1/2} \cos \frac{2j\pi}{M} \right] \right\}.  \label{eq52}
\end{align}
\end{widetext}
In this expression $a_{{\rm{sta}},i}$ and $d_{{\rm{sta}},i}$ are determined by Eq. (\ref{eq51}) with $\alpha = \alpha_i$, and $\left( a_i d_i - b_i c_i \right)_A$ and $\left( a_i d_i - b_i c_i \right)_B$ are determined by Eq. (\ref{eq35}). Therefore, we have shown that $Z_{{\rm{sta}},\left( \text{\romannumeral2} \right)}$ also has an expression of a double product form.

\subsubsection{Thermodynamic limit}
We rearrange Eq. (\ref{eq52}) as we have done in the uniform case
\begin{widetext}
\begin{align}
&Z_{{\rm{sta}},\left( \text{\romannumeral2} \right)} = 2\prod\limits_{i = 1}^{N} \left\{ \left[ \left( \omega _{1A}^2 + \omega _{4A}^2 \right) \omega _{5B} \omega _{7B} + \left( \omega _{2B}^2 + \omega _{3B}^2 \right) \omega _{5A} \omega _{7A} - 2\left( \omega _{1A} \omega _{4A} \omega _{5B} \omega _{7B} - \omega _{5A} \omega _{7A} \omega _{2B} \omega _{3B} \right) \cos {\alpha _i} \right] \right.   \nonumber \\
&~~~~~~~~~~~~~~~~~~~~~\left. \times \prod\limits_{j = 1}^{M/2 - 1} \left[ \left( a_{{\rm{sta}},i} + d_{{\rm{sta}},i} \right)^2 - 4\left( a_i d_i - b_i c_i \right)_A \left( a_i d_i - b_i c_i \right)_B \cos^2 \frac{2j\pi}{M} \right]^{1/2} \right\}.  \label{eq53}
\end{align}
The solution in the thermodynamic limit $M,N \to \infty$ can be directly obtained from Eq. (\ref{eq53}):
\begin{align}
\mathop {\lim }\limits_{M,N \to \infty} \frac{1}{2MN} \ln Z_{{\rm{sta}},\left( \text{\romannumeral2} \right)} = \frac{1}{32 \pi ^2} \int_0^{2\pi } d\theta \int_0^{2\pi} d\phi \ln \left[ \left( a_{\rm{sta}} + d_{\rm{sta}} \right)^2 - 4\left( ad - bc \right)_A \left( ad - bc \right)_B \cos^2 \theta \right],   \label{eq54}
\end{align}
where $a_{\rm{sta}}$ and $d_{\rm{sta}}$ are determined by Eq. (\ref{eq51}) with $\alpha = \phi$, and $\left( ad - bc \right)_A$ and $\left( ad - bc \right)_B$ are determined by Eq. (\ref{eq35}) with $\frac{\left( 2i - 1 \right)\pi}{2N}$ replaced by $\phi$.
\end{widetext}

In the end of this section, we outline the main result. For the row-by-row staggered free-fermion model on an $M\times2N$ square lattice, our desired BCs are the same as those in the uniform case, i.e., BCs $\left( \text{\romannumeral2} \right)$. Under the BCs $\left( \text{\romannumeral2} \right)$ the partition function has an expression of a double product form, as shown in Eq. (\ref{eq52}) or Eq. (\ref{eq53}).

\section{B-K boundary conditions of the square lattice Ising model}   \label{squ}
For a square lattice of $M$ rows and $2N$ columns, the B-K BCs \cite{RN410} for the Ising model are as follows: periodic boundary conditions in the $N$ direction, $2N$ ``$+$'' spins on the upper edge (the $0$th row), and $2N$ alternating spins ``$+-\cdots+-$'' on the lower edge [the $(M+1)$th row].

\begin{figure} 
\includegraphics{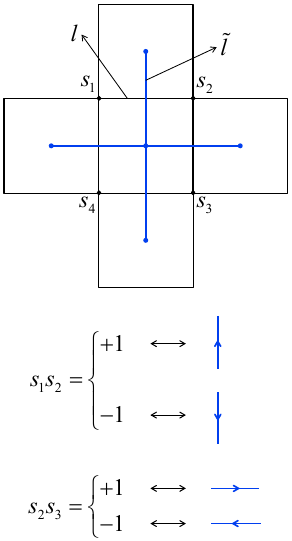}
\caption{Mapping between the arrow configurations of the even eight-vertex model and the spin configurations of the Ising model. The sites and edges of the even eight-vertex model are marked in blue, while those of the Ising model are marked in black.} \label{fig5}
\end{figure}

Reference \cite{RN65} introduced a mapping between the arrow configurations of the even eight-vertex model and the spin configurations of Ising model. Now this mapping method is directly employed to transform the BCs $\left( \text{\romannumeral2} \right)$ into the B-K BCs of the square lattice Ising model. We briefly describe this method as follows. We put a site in the center of each square of the spin lattice, and these sites form a new square lattice, i.e., the dual lattice of the original one \cite{RN236, RN286, RN331, RN284}. Each edge $\tilde l$ of the new square lattice crosses an edge $l$ of the spin lattice. Then we draw an arrow on $\tilde l$ with the correspondence between the arrow configuration and the product of two spins connected by $l$, as shown in Fig. \ref{fig5}. It is easy to verify that, we obtain an even eight-vertex model on the new square lattice by this mapping. Vice versa, given any configuration of an even eight-vertex model we can generate the corresponding spin states.

\begin{figure} 
\includegraphics{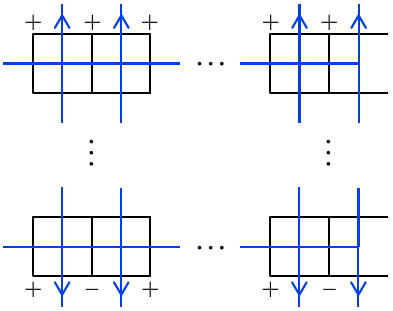}
\caption{The correspondence between the BCs $\left( \text{\romannumeral2} \right)$ of the even eight-vertex model and the B-K BCs of the Ising model.} \label{fig6}
\end{figure}

Now we can indicate the relation between the BCs $\left( \text{\romannumeral2} \right)$ of the even eight-vertex model and the B-K BCs of the square lattice Ising model clearly. First the spins on the upper edge are set as ``$+ \cdots +$''. As shown in Fig. \ref{fig6}, given a certain configuration of the $\left(M+1\right) \times 2N$ even eight-vertex model under the BCs $\left( \text{\romannumeral2} \right)$, we can obtain an $M \times 2N$ spin state. The set of configurations under the BCs $\left( \text{\romannumeral2} \right)$ is mapped into two sets of spin states, i.e., the spins on the lower edge are ``$+-\cdots+-$'' in one set and ``$-+\cdots-+$'' in the other. That is to say, BCs $\left( \text{\romannumeral2} \right)$ correspond to two sets of BCs of the Ising model. Noticing the periodic boundary conditions in the $N$ direction, these two sets of BCs are equivalent. Therefore, BCs $\left( \text{\romannumeral2} \right)$ ($M+1$ rows) are transformed into the B-K BCs ($M$ rows). When the Boltzmann factors of the Ising model are appropriately translated into the vertex weights of the even eight-vertex model, the partition functions can be easily related by
\begin{equation}
Z_{{\rm B}\textit{-}{\rm K}} = \frac{1}{2} Z_{\left( \text{\romannumeral2} \right)}~.   \label{eq55}
\end{equation}

\subsection{Zero-field case}   \label{squ-a}
The Hamiltonian is given in Eq. (\ref{eq1}) with $H_{\rm{ex}}=0$. The square lattice Ising model in this case can be mapped into the uniform free-fermion model, with the vertex weights determined by
\begin{align}
\omega \left( s_1 s_2, s_2 s_3, s_3 s_4, s_1 s_4 \right) = e^{ - \frac{1}{2}\beta J\left( s_1 s_2 + s_2 s_3 + s_3 s_4 + s_1 s_4 \right)}.    \label{eq56}
\end{align}
Here $\left\{ s_1, s_2, s_3, s_4 \right\}$ are the four spins around this vertex site, and four products $\left\{ s_1 s_2, s_2 s_3, s_3 s_4, s_1 s_4 \right\}$ determine the vertex configuration (see Fig. \ref{fig5}).  Note that the energy $2NJ$ from the interactions between the upper edge spins, and $-2NJ$ from the interactions between the lower edge spins, are not included in the Hamiltonian. Equation (\ref{eq56}) takes into account half of both parts, so that the Hamiltonian conserves and the partition function is correct. Now all the eight vertex weights can be listed according to the vertex configurations in Fig. \ref{fig1}: 
\begin{align}
&{\omega _1} = \omega \left( { + , + , + , + } \right) = {e^{ - 2\beta J}},  \nonumber \\ 
&{\omega _2} = \omega \left( { - , - , - , - } \right) = {e^{2\beta J}},  \nonumber \\
&{\omega _3} = \omega \left( { - , + , - , + } \right) = 1,  \nonumber \\
&{\omega _4} =  \cdots  = {\omega _8} = 1.   \label{eq57}
\end{align}
Obviously the free-fermion condition is satisfied. Therefore, the mapping leads to an $\left(M+1\right) \times 2N$ free-fermion model under the BCs $\left( \text{\romannumeral2} \right)$. Using Eqs. (\ref{eq55}) and (\ref{eq34}), we immediately obtain the solution
\begin{align}
&Z_{{\rm B}\textit{-}{\rm K}, {\rm squ}} = 4^{MN} \prod\limits_{i = 1}^N \prod\limits_{j = 1}^{M} \left[ \cosh^2 \left(2 \beta J \right) + \sinh \left(2 \beta J \right)  \right.   \nonumber \\
&~~~~~~~~~~~~~~~~~~~\left. \times \left( \cos \frac{\left( 2i - 1 \right)\pi}{2N} + \cos \frac{j\pi}{M+1} \right) \right].   \label{eq58} 
\end{align}
Note that this expression holds for both cases that $J>0$ and $J<0$. This well-known solution has been studied in various literatures \cite{RN410, RN474, RN455, RN475}. The partition function in the thermodynamic limit is yielded \cite{RN72, RN267, RN74, RN207, RN76, RN265, RN269, RN209}
\begin{align}
&\mathop {\lim }\limits_{M,N \to \infty} \frac{1}{2MN} \ln Z_{{\rm B}\textit{-}{\rm K}, {\rm squ}} = \ln 2 + \frac{1}{8 \pi ^2}\int_0^{2\pi } d\theta \int_0^{2\pi } d\phi  \nonumber \\ 
&~~~~~~ \ln \left[ {\cosh}^2 \left( 2\beta J \right) + \sinh \left( 2\beta J \right) \left( \cos \phi + \cos \theta \right) \right].   \label{eq59}
\end{align} 

The Fisher zeros in the zero field can then be conveniently calculated from Eq. (\ref{eq58}). Given $i$ and $j$, the zeros in the variable $\bar z = \sinh \left(2\beta J\right)$ are $e^{\pm i \eta_{ij}}$ with $\eta _{ij} \in \left[ 0, \pi \right]$ given by
\begin{align}
\cos {\eta_{ij}} =  - \frac{1}{2}\left( \cos \frac{\left( 2i - 1 \right)\pi}{2N} + \cos \frac{j\pi}{M+1} \right).   \label{eq60}
\end{align}
The Fisher zeros in the complex $\bar z$ plane lie on the unit circle $\left| \bar z \right| = 1$. Then we consider the variable $z=e^{2\beta J}$, which is usually used in the low-temperature series expansion. Given $i$ and $j$, four zeros in the variable $z$ can be calculated, thus the $4MN$ zeros are explicitly determined. When $0 \le \eta _{ij} \le \frac{\pi }{2}$, the zeros can be expressed as 
\begin{align}
z_{ij,k} = 1 + \sqrt 2 e^{i \theta_{ij,k}},~k = 1, \cdots ,4     \label{eq61}
\end{align}
with
\begin{align}
&\theta _{ij,1} = \frac{1}{2} \eta_{ij} + \arccos \left( \sqrt{\cos \eta_{ij}} \right) ,  \nonumber \\
&\theta _{ij,2} = \frac{1}{2} \eta_{ij} - \arccos \left( \sqrt{\cos \eta_{ij}} \right) + \pi ,  \nonumber \\
&\theta _{ij,3} = -\frac{1}{2} \eta_{ij} - \arccos \left( \sqrt{\cos \eta_{ij}} \right) + 2\pi ,  \nonumber \\
&\theta _{ij,4} = -\frac{1}{2} \eta_{ij} + \arccos \left( \sqrt{\cos \eta_{ij}} \right) + \pi .   \label{eq62}
\end{align}
While $\frac{\pi}{2} < \eta_{ij} \le \pi$, the zeros are
\begin{align}
z_{ij,k} = -1 + \sqrt 2 e^{i \theta_{ij,k}},~k = 1, \cdots ,4     \label{eq63}
\end{align}
with
\begin{align}
&\theta _{ij,1} = \frac{1}{2} \eta_{ij} + \arccos \left( \sqrt{-\cos \eta_{ij}} \right) - \frac{\pi}{2} ,  \nonumber \\
&\theta _{ij,2} = \frac{1}{2} \eta_{ij} - \arccos \left( \sqrt{-\cos \eta_{ij}} \right) + \frac{\pi}{2} ,  \nonumber \\
&\theta _{ij,3} = -\frac{1}{2} \eta_{ij} - \arccos \left( \sqrt{-\cos \eta_{ij}} \right) + \frac{5\pi}{2} ,  \nonumber \\
&\theta _{ij,4} = -\frac{1}{2} \eta_{ij} + \arccos \left( \sqrt{-\cos \eta_{ij}} \right) + \frac{3\pi}{2} .   \label{eq64}
\end{align}
We see that, the Fisher zeros in the complex $z$ plane lie on two circles
\begin{align}
\left| {z \pm 1} \right| = \sqrt 2~,   \label{eq65}
\end{align}
for any finite lattice under the B-K BCs.

\begin{figure} 
\includegraphics{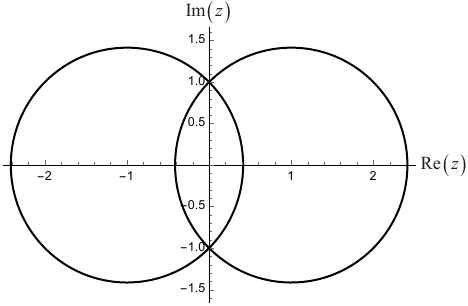}
\caption{The Fisher loci in the complex $z$ plane of the square lattice Ising model in the zero field.} \label{fig7}
\end{figure}

As the system approaches the thermodynamic limit $M,N \to \infty$, the number of zeros increases to infinity, and the accumulation points of the zeros, i.e., the Fisher loci, form these two circles. This is consistent with the original finding of Fisher \cite{RN308}. Figure \ref{fig7} shows these two circles. In the thermodynamic limit, the distribution of zeros on the Fisher loci $\pm 1 + \sqrt 2 e^{i\theta}\left(0 \le \theta < 2\pi \right)$ can be described by a density function. Lu and Wu derived the density functions in the complex $\bar z$ plane and $\tanh \left( - \beta J \right)$ plane \cite{RN411}, and the transformation into the variable $z$ is straightforward. We give the explicit expression of this density function \cite{RN529} 
\begin{align}
&g_{\rm{l}}\left( \theta  \right) = \frac{{2\left| {\sin \theta } \right|\left( {\sqrt 2  - \cos \theta } \right)\left| {1 - \sqrt 2 \cos \theta } \right|}}{{{\pi ^2}{{\left( {3 - 2\sqrt 2 \cos \theta } \right)}^2}}}   \nonumber \\
&~~~~~~~~~~ \times K\left( {\frac{{2\sin \theta \left( {\sqrt 2  - \cos \theta } \right)}}{{3 - 2\sqrt 2 \cos \theta }}} \right),   \nonumber \\
&g_{\rm{r}}\left( \theta  \right) = g_{\rm{l}}\left( \left| \pi - \theta \right| \right),   \label{eq66}
\end{align}
where $K\left( k \right) = \int_0^{\pi/2} \frac{1}{\sqrt{1 - {k^2}{\sin^2}t}}dt$ is the complete elliptic integral of the first kind, l and r refer to the circles on the left and right, respectively. The Fisher loci cut the physical domain of $z$, i.e., the interval $\left(0, 1\right)$ [or $\left(1, +\infty \right)$] on the positive real axis, at the critical point $z = \sqrt{2}-1$ (or $z = \sqrt{2}+1$). It is clear to verify from Eq. (\ref{eq66}) that, the density function near the critical point (when $\left| \theta \right| \to 0$) is of the order $O\left( \left| \theta \right| \right)$. This leads to a divergence of the specific heat, i.e., a second-order phase transition at the critical point in the thermodynamic limit \cite{RN308, RN411, RN529}. 

\subsection{Imaginary-field case}   \label{squ-b}
\begin{figure} 
\includegraphics{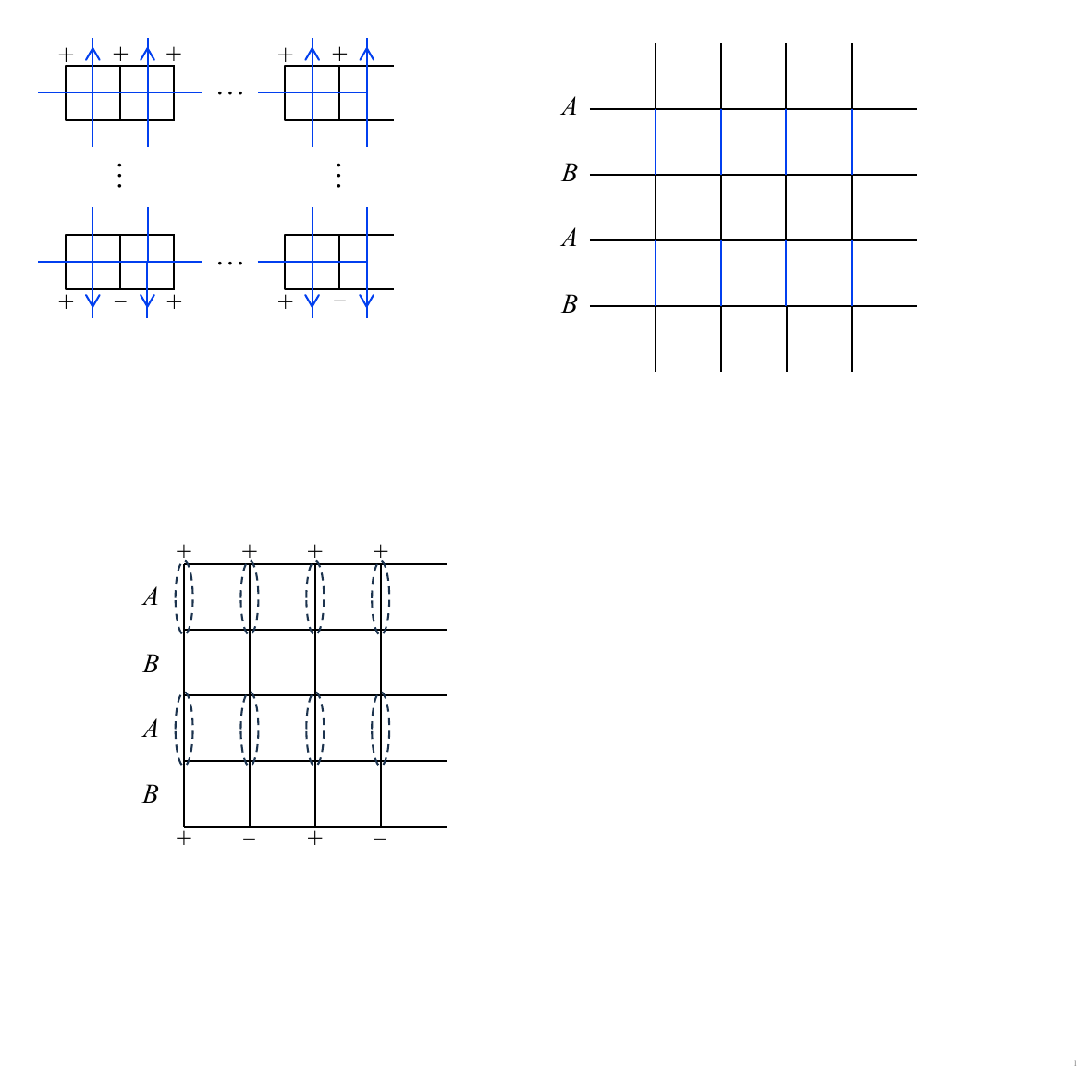}
\caption{Splitting of $\prod\nolimits_i {s_i}$ into certain pairs of product $s_i s_j$ covering all spin sites. An example of $3\times 4$ square lattice is shown, with the pairs marked with dash lines.} \label{fig8}
\end{figure}

In this case, the effect of the imaginary field $H_{\rm{ex}}=i(\pi/2)k_BT$ is deduced as \cite{RN51, RN274, RN558, RN529}
\begin{align}
e^{\beta H_{\rm{ex}}\sum\limits_{i = 1}^{2MN} {s_i}} = i^{\sum\limits_{i = 1}^{2MN} {s_i}} = i^{2MN} \prod\limits_{i = 1}^{2MN} {s_i}~.  \label{eq67}
\end{align}
In the last step we use the identity $i^{s_i} = i\times s_i$. Omitting the constant $ i^{2MN}$, the contribution of a certain spin state $\left\{ s_i \right\}$ to the partition function includes the Boltzmann factor $e^{-\beta J \sum\nolimits_{\left\langle {ij} \right\rangle}{s_i s_j}}$ and the product of all spins $\prod\nolimits_i {s_i}$. Following Refs. \cite{RN70} and \cite{RN274}, we split $\prod\nolimits_i {s_i}$ into certain pairs of product $s_i s_j$ covering all the sites. Taking the $3\times 4$ lattice as an example, the splitting method is shown in Fig. \ref{fig8}. Note that the $2N$ ``$+$'' spins on the upper edge are included in the product, which does not affect the correctness. From Fig. \ref{fig8} we can clearly see that this case is mapped into a row-by-row staggered free-fermion model, by defining the vertex weights on the sublattice $A$ as
\begin{align}
&\omega_A \left( s_1 s_2, s_2 s_3, s_3 s_4, s_1 s_4 \right) = \left(s_1 s_4\right)   \nonumber \\
&~~~~~~~~~~~~~~~~~~~~~~~~~\times e^{ - \frac{1}{2}\beta J\left( s_1 s_2 + s_2 s_3 + s_3 s_4 + s_1 s_4 \right)},    \label{eq68}
\end{align}
and those on the sublattice $B$ as Eq. (\ref{eq56}). We calculate all the vertex weights:
\begin{align}
&\omega _{1A} = \omega_A \left( +, +, +, + \right) = e^{ - 2\beta J},  \nonumber \\ 
&\omega _{2A} = \omega_A \left( -, -, -, - \right) = -e^{2\beta J},  \nonumber \\
&\omega _{3A} = \omega_A \left( -, +, -, + \right) = 1,  \nonumber \\
&\omega _{4A} = \omega_A \left( +, -, +, - \right) = -1,  \nonumber \\
&\omega _{5A} = \omega _{8A} = 1,  \nonumber \\
&\omega _{6A} = \omega _{7A} = -1,  \label{eq69}
\end{align}
and $\left\{ \omega_B \right\}$ are listed in Eq. (\ref{eq57}).

Now we can give the solution of this case when $M$ is odd, as the row-by-row staggered free-fermion model under the BCs $\left( \text{\romannumeral2} \right)$ with an even number of rows has been exactly solved. Substituting the vertex weights into Eq. (\ref{eq52}) yields 
\begin{align}
&~~~~Z_{{\rm B}\textit{-}{\rm K}, {\rm squ}, i\frac{\pi}{2}}  \nonumber \\
& = \prod\limits_{i = 1}^{N} \left\{ \left[ -4\sinh \left(2\beta J\right) \left( \cosh \left(2\beta J\right) + \cos \frac{\left(2i-1\right) \pi}{2N} \right) \right] \right.  \nonumber \\
&~~~~~~~~~~ \times \prod\limits_{j = 1}^{(M-1)/2} \left[ 8\sinh^2 \left(2\beta J\right) \left( 2\cosh^2 \left(2\beta J\right)  \right. \right.  \nonumber \\
&~~~~~~~~~~~~~~~~~~ \left. \left. \left.  - \cos \frac{\left(2i-1\right) \pi}{N} - \cos \frac{2j\pi}{M+1} \right) \right] \right\}.  \label{eq70}
\end{align}
The solution when $M$ is even has been given in Ref. \cite{RN432}. The partition function in the thermodynamic limit is directly obtained \cite{RN57, RN67, RN71, RN68, RN69, RN70, RN66, RN274}
\begin{align}
&\mathop {\lim }\limits_{M,N \to \infty} \frac{1}{2MN} \ln Z_{{\rm B}\textit{-}{\rm K}, {\rm squ}, i\frac{\pi}{2}} = \frac{1}{16 \pi ^2}\int_0^{2\pi } d\theta \int_0^{2\pi } d\phi  \ln \left[ 16  \right.  \nonumber \\ 
& \left. \times \sinh^2 \left( 2\beta J \right) \cosh^2 \left( 2\beta J \right) - 8\sinh^2 \left( 2\beta J \right) \left( \cos \phi + \cos \theta \right) \right].   \label{eq71}
\end{align} 

\begin{figure} 
\includegraphics{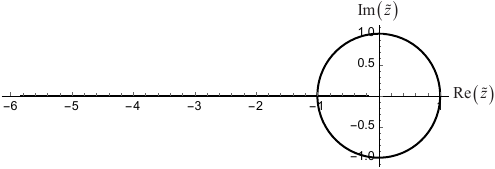}
\caption{The Fsiher loci in the complex $\tilde z$ plane of the square lattice Ising model in the imaginary field.} \label{fig9}
\end{figure}

It is convenient to calculate the Fisher zeros in the variable $\tilde z = e^{4\beta J}$. First the partition function in Eq. (\ref{eq70}) is expressed as
\begin{align}
\!\!\!\!&~~~~Z_{{\rm B}\textit{-}{\rm K}, {\rm squ}, i\frac{\pi}{2}}   \nonumber \\
\!\!\!\!& = \left( \frac{1-\tilde z}{\tilde z} \right)^{MN} \prod\limits_{i = 1}^{N} \left\{ \left[ 1 + \tilde z + 2\tilde z^{1/2} \cos \frac{\left(2i-1\right) \pi}{2N} \right]  \right.   \nonumber \\
\!\!\!\!&\left. \times \prod\limits_{j = 1}^{(M-1)/2} \left[ 1 + \tilde z^2 + 2\tilde z \left( 1 - \cos \frac{\left(2i-1\right) \pi}{N} - \cos \frac{2j\pi}{M+1} \right) \right] \right\}   \nonumber \\ 
\!\!\!\!&= \left( \frac{1-\tilde z}{\tilde z} \right)^{MN} \left(1 + \tilde z^N \right) \prod\limits_{i = 1}^{N} \prod\limits_{j = 1}^{(M-1)/2} \left[ 1 + \tilde z^2 + 2\tilde z \left( 1  \right. \right.   \nonumber \\
\!\!\!\!&~~~~~~~~~~~~~~~~~~~\left. \left. - \cos \frac{\left(2i-1\right) \pi}{N} - \cos \frac{2j\pi}{M+1} \right) \right].   \label{eq72}
\end{align}
In the last step we have made use of
\begin{align}
\gamma^N + \gamma^{-N} = \prod\limits_{i = 1}^N \left[ \gamma + \gamma^{-1} + 2\cos \frac{\left( 2i - 1 \right)\pi}{2N} \right].   \label{eq73}
\end{align}
Equation (\ref{eq72}) demonstrates that the $2MN$ zeros in the complex $\tilde z$ plane include the root 1 of multiplicity $MN$, $N$ roots on the unit circle associated with $1 + \tilde z^N$, and $\left(M-1\right)N$ roots $\left\{ \tilde z_{ij} \right\}$ from the term
\begin{align*}
\!\!\! \prod\limits_{i = 1}^{N} \prod\limits_{j = 1}^{(M-1)/2} \left[ 1 + \tilde z^2 + 2\tilde z \left( 1 - \cos \frac{\left(2i-1\right) \pi}{N} - \cos \frac{2j\pi}{M+1} \right) \right].
\end{align*}
The loci of zeros $\left\{ \tilde z_{ij} \right\}$ are similar to that in the case where $M$ is even, as analysed in Ref. \cite{RN432}. The zeros $\left\{ \tilde z_{ij} \right\}$ lie on the unit circle
\begin{subequations}
\label{eq74}
\begin{equation}
\left| \tilde z \right| = 1,~{\rm{for}}~ -1 \le 1 - \cos \frac{\left(2i-1\right) \pi}{N} - \cos \frac{2j\pi}{M+1} \le 1,   \label{eq74a}
\end{equation}
and on the line segment
\begin{align}
& -3-2\sqrt 2 \le \tilde z \le -3+2\sqrt 2 ,~{\rm{for}}~ 1 < 1 - \cos \frac{\left(2i-1\right) \pi}{N}   \nonumber \\
&~~~~~~~~~~~~~~~~~~~~~~~~~~~~~~~~~~~~~ - \cos \frac{2j\pi}{M+1} \le 3.   \label{eq74b}
\end{align}
\end{subequations}
We see that the Fisher loci in the complex $\tilde z$ plane in the thermodynamic limit consist of the unit circle and the line segment \cite{RN430, RN505, RN432}, as shown in Fig. \ref{fig9}. The density function of Fisher zeros in the thermodynamic limit depends on the distribution of $\left\{ \tilde z_{ij} \right\}$. The explicit expression is \cite{RN411}:
\begin{subequations}
\label{eq75}
\begin{equation}
g_{\rm{c}} \left( \alpha \right) = \frac{\left| \sin \alpha \right|}{2{\pi ^2}}K\left( \sqrt{\frac{\left( 3 + \cos \alpha \right)\left( 1 - \cos \alpha \right)}{4}} \right)   \label{eq75a}
\end{equation}
on the unit circle $e^{i\alpha}\left( 0 \le \alpha < 2\pi \right)$, and
\begin{align}
\!\! g_{\rm{li}} \left( \lambda \right) = \frac{\left| \sinh \lambda \right|}{2{\pi^2}}K\left( \sqrt{\frac{\left( 3 - \cosh \lambda \right)\left( 1 + \cosh \lambda \right)}{4}} \right)  \label{eq75b}
\end{align}
on the line segment $- e^\lambda \left[ \ln \left( 3 - 2\sqrt 2 \right) \le \lambda \le \ln \left( 3 + 2\sqrt 2 \right) \right]$.
\end{subequations}
The Fisher loci do not cut the interval $\left(0,1 \right)$ [or $\left(1, +\infty \right)$] on the positive real axis, thus the system in this case does not have a physical phase transition.

\section{B-K type boundary conditions of the triangular lattice Ising model}   \label{tri}
\begin{figure} 
\includegraphics{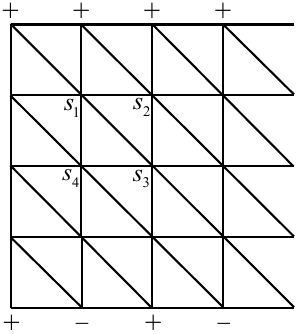}
\caption{The B-K type BCs of the triangular lattice Ising model.} \label{fig10}
\end{figure}

The B-K BCs of the square lattice Ising model can be extended to the triangular and honeycomb lattices. In this section we introduce the B-K type BCs for the triangular lattice model, as shown in Fig. \ref{fig10}. The triangular lattice is easily represented by a ``square'' structure, with $M$ rows and $2N$ columns. Then it is straightforward to propose the B-K type BCs: periodic boundary conditions in the $N$ direction, fixed spins ``$+\cdots+$'' on the upper edge, and fixed spins ``$+-\cdots+-$'' on the lower edge. 

We employ the same mapping method as that shown in Fig. \ref{fig5}, to construct an even eight-vertex model under the BCs $\left( \text{\romannumeral2} \right)$ with $M+1$ rows and $2N$ columns. As indicated in the case of the square lattice Ising model, the relation between the partition functions---Eq. (\ref{eq55})---still holds, if the vertex weights can be appropriately defined using the Boltzmann factors of the Ising model.

\subsection{Zero-field case}   \label{tri-a}
This case can be mapped into the uniform free-fermion model, whose vertex weights are given by
\begin{align}
\!\! \omega \left( s_1 s_2, s_2 s_3, s_3 s_4, s_1 s_4 \right) = e^{-\beta J\left[ \frac{1}{2}\left( s_1 s_2 + s_2 s_3 + s_3 s_4 + s_1 s_4 \right) + s_1 s_3 \right]}.    \label{eq76}
\end{align}
The determination of the vertex weights is very similar to that for the square lattice Ising model, except for the interaction $Js_1 s_3 = J \left(s_1 s_2\right)\left(s_2 s_3\right)$ (see Fig. \ref{fig10}). All vertex weights are then calculated
\begin{align}
&\omega _1 = e^{-3\beta J},~ \omega _2 = e^{\beta J},   \nonumber \\
&\omega _3 = \cdots = \omega_6 = e^{\beta J},~\omega _7 = \omega _8 = e^{-\beta J}.   \label{eq77}
\end{align}
Using Eq. (\ref{eq34}) with the vertex weights in Eq. (\ref{eq77}), the solution is obtained
\begin{align}
&Z_{{\rm B}\textit{-}{\rm K}, {\rm tri}} = \prod\limits_{i = 1}^N \prod\limits_{j = 1}^{M} \left[ e^{-6\beta J} + 3e^{2\beta J} + 4\sinh \left(2\beta J \right)  \right.   \nonumber \\
&~~~~~\left. \times \left( \cos \frac{\left( 2i - 1 \right)\pi}{2N} + 2\cos \frac{\left( 2i - 1 \right)\pi}{4N} \cos \frac{j\pi}{M+1} \right) \right].   \label{eq78} 
\end{align}
The partition function in the thermodynamic limit is directly yielded
\begin{align*}
&\mathop {\lim }\limits_{M,N \to \infty} \frac{1}{2MN} \ln Z_{{\rm B}\textit{-}{\rm K}, {\rm tri}} = \frac{1}{\pi ^2}\int_0^{\pi} d\theta \int_0^{\pi/2} d\phi  \ln \left[ e^{-6\beta J}  \right.   \nonumber \\ 
&~~~~~~~~~~~ \left. + 3e^{2\beta J} + 4\sinh \left(2\beta J \right) \left( \cos 2\phi + 2\cos \phi \cos \theta \right) \right].  
\end{align*}
By performing variable transformation $\tilde \theta = \theta + \phi,~\tilde \phi = \theta - \phi$ and doing some simple algebra, we achieve the well-known expression \cite{RN81, RN299, RN223, RN558}
\begin{align}
&\mathop {\lim }\limits_{M,N \to \infty} \frac{1}{2MN} \ln Z_{{\rm B}\textit{-}{\rm K}, {\rm tri}} = \frac{1}{8\pi ^2}\int_0^{2\pi} d{\tilde \theta} \int_0^{2\pi} d{\tilde \phi}  \ln \left[ e^{-6\beta J}  \right.   \nonumber \\ 
&~~~~~ \left. + 3e^{2\beta J} + 4\sinh \left(2\beta J \right) \left( \cos{\tilde \theta} + \cos{\tilde \phi} + \cos ( \tilde \theta - \tilde \phi ) \right) \right].    \label{eq79}
\end{align}

\begin{figure} 
\includegraphics{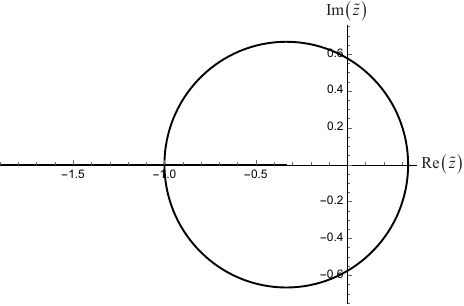}
\caption{The Fisher loci in the complex $\tilde z$ plane of the triangular lattice Ising model in the zero field.} \label{fig11}
\end{figure}

We use the variable $\tilde z = e^{4\beta J}$ to solve the Fisher zeros. The solution in Eq. (\ref{eq78}) has the form
\begin{align}
Z_{{\rm B}\textit{-}{\rm K}, {\rm tri}} & = \tilde z^{-3MN/2} \prod\limits_{i = 1}^N \prod\limits_{j = 1}^{M} \left[ 1 + 3\tilde z^2 + 2\tilde z\left(\tilde z -1 \right)  \right.   \nonumber \\
&\left. \times \left( \cos \frac{\left( 2i - 1 \right)\pi}{2N} + 2\cos \frac{\left( 2i - 1 \right)\pi}{4N} \cos \frac{j\pi}{M+1} \right) \right]   \nonumber \\
& \equiv \tilde z^{-3MN/2} \prod\limits_{i = 1}^N \prod\limits_{j = 1}^{M} \left[ 1 + 3\tilde z^2 + 2\tilde z\left(\tilde z -1 \right) \left( \cos_{ij} \right) \right],    \label{eq80} 
\end{align}
where the term $\left( \cos_{ij} \right)$ denotes the part of the cosine functions. The $2MN$ Fisher zeros can then be solved from $1 + 3\tilde z^2 + 2\tilde z\left(\tilde z -1 \right) \left( \cos_{ij} \right)$
\begin{equation}
\tilde z_{ij,1,2} = \frac{\left( \cos_{ij} \right) \pm \sqrt{\left[ \left(\cos_{ij}\right) - 3 \right] \left[ \left(\cos_{ij}\right) + 1 \right]}} {3 + 2\left(\cos_{ij}\right)}~.   \label{eq81}
\end{equation}
It is trivial to verify that $-\frac{3}{2} < \left( \cos_{ij} \right) < 3$. Therefore, when $-\frac{3}{2} < \left( \cos_{ij} \right) \le -1$, $\tilde z_{ij,1}$ and $\tilde z_{ij,2}$ are negative real numbers in the respective domains
\begin{equation}
\tilde z_{ij,1} \in \left[ -1, -\frac{1}{3} \right),~\tilde z_{ij,2} \in \left( -\infty, -1 \right].   \label{eq82}
\end{equation}
When $-1 < \left( \cos_{ij} \right) < 3$, we have
\begin{equation}
\tilde z_{ij,1,2} = \frac{\left( \cos_{ij} \right) \pm i\sqrt{\left[ 3 - \left(\cos_{ij}\right) \right] \left[ 1 + \left(\cos_{ij}\right) \right]}} {3 + 2\left(\cos_{ij}\right)}~,   \label{eq83}
\end{equation}
and one can examine that $\tilde z_{ij,1}$ and $\tilde z_{ij,2}$ can be expressed as $\tilde z_{ij,k} = -\frac{1}{3} + \frac{2}{3}e^{i\theta_{ij,k}}$ with
\begin{equation}
\theta_{ij,1} \in \left( 0, \pi \right),~\theta_{ij,2} = -\theta_{ij,1}+2\pi \in \left( \pi, 2\pi \right).   \label{eq84}
\end{equation}
The Fisher loci in the thermodynamic limit consist of the line segment $\left( -\infty, -\frac{1}{3} \right]$ and the circle $-\frac{1}{3} + \frac{2}{3}e^{i\theta} \left(0 \le \theta < 2\pi \right)$, which can be seen in Fig. \ref{fig11}. The exact Fisher loci had been reported in Refs. \cite{RN427, RN458, RN436}, by simply setting the argument of the logarithm in the solution in the thermodynamic limit equal to zero, which was termed as ``handwaving'' in Ref. \cite{RN427}. Here the use of the B-K type BCs makes the determination rigorous, as the Fisher zeros of any finite lattice under the B-K type BCs can be calculated exactly, and found to be precisely on certain loci. 

The Fisher loci cut the positive real axis at $\tilde z = \frac{1}{3}$, which is the ferromagnetic critical point. This elucidates that, only the ferromagnetic model, i.e., the $J<0$ case, exhibits a physical phase transition. The density function of Fisher zeros in the thermodynamic limit had also been derived \cite{RN411}, and we quote it here:
\begin{align}
g_{\rm{c}}\left( \theta \right) = \frac{3^{3/2} \left| \sin \theta \right|}{\pi^2 \left( 5 - 4\cos \theta \right)^{7/4}} K\left[ k\left(\theta\right) \right]   \label{eq85}
\end{align}
on the circle $-\frac{1}{3} + \frac{2}{3}e^{i\theta}$ with
\begin{align}
&k\left(\theta \right) = \frac{1}{4} \left[\frac{3}{E\left(\theta \right)}-1\right]^{1/2} \left[1 + E\left(\theta \right) \right]^{3/2}~,   \nonumber \\
&E\left(\theta \right) = \frac{3}{\left(5-4\cos\theta \right)^{1/2}}~,   \label{eq86}
\end{align}
and 
\begin{align}
g_{\rm{l}}\left( u \right) = \frac{\left| 1+u \right| \left(1-3u\right)}{4\pi^2 u^2 \left( 1-u \right)^2 k\left(u\right) F^{1/2}\left(u\right)} K\left[ \frac{1}{k\left(u\right)} \right]   \label{eq87}
\end{align}
on the line segment $-\infty < u \le -\frac{1}{3}$ with
\begin{align}
&k\left(u \right) = \frac{1}{4} \left[\frac{3}{F\left(u \right)}-1\right]^{1/2} \left[1 + F\left(u \right) \right]^{3/2}~,   \nonumber \\
&F\left(u \right) = \left[\frac{1+3u}{u \left(1-u \right)}\right]^{1/2}~.   \label{eq88}
\end{align}
The density function $g_{\rm{c}}\left( \theta \right)$ near the critical point (when $\left| \theta \right| \to 0$) is of the order $O\left( \left| \theta \right| \right)$. This verifies a second-order phase transition at the ferromagnetic critical point.

\subsection{Imaginary-field case}   \label{tri-b}
We still use Eq. (\ref{eq67}) to deal with the effect of the imaginary field, and split the product $\prod\nolimits_i {s_i}$ in the way shown in Fig. \ref{fig8}, when $M$ is odd. Like in the case of the square lattice model in the imaginary field, we map this case into a row-by-row staggered free-fermion model under the BCs $\left( \text{\romannumeral2} \right)$, with $\left\{\omega_A\right\}$ defined as
\begin{align}
&\omega_A \left( s_1 s_2, s_2 s_3, s_3 s_4, s_1 s_4 \right) = \left(s_1 s_4\right)   \nonumber \\
&~~~~~~~~~~~~~~~~\times e^{-\beta J\left[ \frac{1}{2}\left( s_1 s_2 + s_2 s_3 + s_3 s_4 + s_1 s_4 \right) + s_1 s_3 \right]}    \label{eq89}
\end{align}
and $\left\{\omega_B\right\}$ determined from Eq. (\ref{eq76}). We then calculate all the vertex weights:
\begin{align}
&\omega _{1A} = e^{-3\beta J},~ \omega _{2A} = -e^{\beta J},~\omega _{3A} = \omega_{5A} = e^{\beta J},   \nonumber \\
&\omega _{4A} = \omega_{6A} = -e^{\beta J},~\omega _{7A} = -e^{-\beta J},~\omega _{8A} = e^{-\beta J},   \label{eq90}
\end{align}
and $\left\{\omega_B\right\}$ are listed in Eq. (\ref{eq77}). Using Eq. (\ref{eq52}) the solution of this case is immediately obtained
\begin{align}
&~~~~Z_{{\rm B}\textit{-}{\rm K}, {\rm tri}, i\frac{\pi}{2}}   \nonumber \\
& = \prod\limits_{i = 1}^{N} \left\{ \left[ -2\sinh \left(2\beta J\right) \left( e^{-4\beta J} + 1 + 2\cos \frac{\left(2i-1\right) \pi}{2N} \right) \right] \right.  \nonumber \\
&~~~~~~ \times \prod\limits_{j = 1}^{(M-1)/2} \left[ e^{-12\beta J} - 4 + 3e^{4\beta J} - 8\sinh^2 \left(2\beta J\right)   \right.  \nonumber \\
&~~~~~ \left. \left. \times \left( \cos \frac{\left(2i-1\right) \pi}{N} + 2\sin \frac{\left(2i-1\right) \pi}{2N} \cos \frac{2j\pi}{M+1} \right) \right] \right\}.  \label{eq91}
\end{align}
It is straightforward to yield the partition function in the thermodynamic limit 
\begin{align*}
\!\!&\mathop {\lim }\limits_{M,N \to \infty} \frac{1}{2MN} \ln Z_{{\rm B}\textit{-}{\rm K}, {\rm tri}, i\frac{\pi}{2}} = \frac{1}{4\pi ^2}\int_0^{\pi} d\theta \int_0^{\pi} d\phi  \ln \left[ e^{-12\beta J}  \right.   \nonumber \\ 
\!\!&~~~~~~~ \left. - 4 + 3e^{4\beta J} - 8\sinh^2 \left(2\beta J\right) \left( \cos 2\phi + 2\sin \phi \cos \theta \right) \right].  
\end{align*}
By performing variable transformation $\tilde \phi = \phi + \theta - \frac{\pi}{2},~\tilde \theta = \phi - \theta + \frac{\pi}{2}$ and doing some simple algebra, we achieve the known expression \cite{RN67, RN274, RN558}
\begin{align}
&\mathop {\lim }\limits_{M,N \to \infty} \frac{1}{2MN} \ln Z_{{\rm B}\textit{-}{\rm K}, {\rm tri}, i\frac{\pi}{2}} = \frac{1}{16\pi ^2}\int_0^{2\pi} d{\tilde \theta} \int_0^{2\pi} d{\tilde \phi}   \nonumber \\ 
&~~\ln \left[ e^{-12\beta J} - 4 + 3e^{4\beta J} - 8\sinh^2 \left(2\beta J\right) \left( \cos{\tilde \phi} - \cos{\tilde \theta}   \right. \right.   \nonumber \\
&~~~~~~~~~~~~~~~~~~~~~~~~~~~~~~~~~~~~~~~~~~\left. \left. + \cos(\tilde \theta + \tilde \phi) \right) \right].    \label{eq92}
\end{align}

\begin{figure} 
\includegraphics{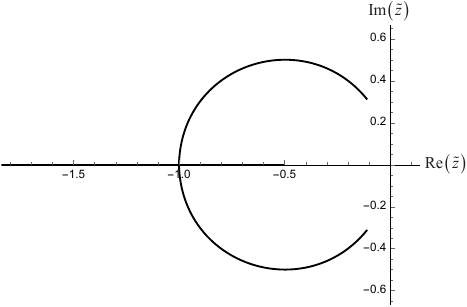}
\caption{The Fisher loci in the complex $\tilde z$ plane of the triangular lattice Ising model in the imaginary field.} \label{fig12}
\end{figure}

We still express the solution in Eq. (\ref{eq91}) in terms of the variable $\tilde z = e^{4\beta J}$
\begin{align}
\!\!&~~~~Z_{{\rm B}\textit{-}{\rm K}, {\rm tri}, i\frac{\pi}{2}}   \nonumber \\
\!\!& = \tilde z^{-3MN/2} \left(1-\tilde z\right)^{MN} \prod\limits_{i = 1}^N \left\{ \left[ 1 + \tilde z \left(1 + 2\cos \frac{\left(2i-1\right) \pi}{2N} \right) \right]  \right.   \nonumber \\
\!\!&~~~~~~~~~~~~~~~\left. \times \prod\limits_{j = 1}^{(M-1)/2}\left[ \left( 1+\tilde z \right)^2 + 4\tilde z^2 \left( \sin_{ij} \right) \right] \right\},    \label{eq93} 
\end{align}
where $\left( \sin_{ij} \right)$ denotes
\begin{align*}
\sin \frac{\left(2i-1\right) \pi}{2N} \left(\sin \frac{\left(2i-1\right) \pi}{2N} - \cos \frac{2j\pi}{M+1} \right)~.
\end{align*}
We can indicate from Eq. (\ref{eq93})  that the $2MN$ zeros in the complex $\tilde z$ plane include the root 1 of multiplicity $MN$, $N$ roots associated with $1 + \tilde z \left(1 + 2\cos \frac{\left(2i-1\right) \pi}{2N} \right)$, and $\left(M-1\right)N$ roots $\left\{ \tilde z_{ij} \right\}$ from the term
\begin{align*}
\prod\limits_{i = 1}^{N} \prod\limits_{j = 1}^{(M-1)/2} \left[ \left( 1+\tilde z \right)^2 + 4\tilde z^2 \left( \sin_{ij} \right) \right].
\end{align*}
The Fisher loci in the thermodynamic limit only depend on the distribution of $\left\{ \tilde z_{ij} \right\}$, which exhibits a non-zero and non-trivial density. $\left\{ \tilde z_{ij} \right\}$ can be directly solved 
\begin{equation}
\tilde z_{ij,1,2} = \frac{-1 \pm 2\sqrt{-\left(\sin_{ij}\right)}} {1 + 4\left(\sin_{ij}\right)}~.   \label{eq94}
\end{equation}
One can verify that $-\frac{1}{4} < \left( \sin_{ij} \right) < 2$. When $-\frac{1}{4} < \left( \sin_{ij} \right) \le 0$, $\tilde z_{ij,1}$ and $\tilde z_{ij,2}$ are negative real numbers in the respective domains
\begin{equation}
\tilde z_{ij,1} \in \left[ -1, -\frac{1}{2} \right),~\tilde z_{ij,2} \in \left( -\infty, -1 \right].   \label{eq95}
\end{equation}
When $0 < \left( \sin_{ij} \right) < 2$, we have
\begin{equation}
\tilde z_{ij,1,2} = \frac{-1 \pm 2i\sqrt{\left(\sin_{ij}\right)}} {1 + 4\left(\sin_{ij}\right)}~.   \label{eq96}
\end{equation}
It can be examined that $\tilde z_{ij,1}$ and $\tilde z_{ij,2}$ in this case can be expressed as $\tilde z_{ij,k} = -\frac{1}{2} + \frac{1}{2}e^{i\theta_{ij,k}}$ with
\begin{align}
&\theta_{ij,1} \in \left( \theta_0, \pi \right),~\theta_{ij,2} = -\theta_{ij,1}+2\pi \in \left( \pi, -\theta_0+2\pi \right),  \nonumber \\
&\theta_0 = \arccos \left(\frac{7}{9}\right).   \label{eq97}
\end{align}
The Fisher loci in the thermodynamic limit consist of the line segment $\left( -\infty, -\frac{1}{2} \right]$ and the arc $-\frac{1}{2} + \frac{1}{2}e^{i\theta} \left(\theta_0 \le \theta \le -\theta_0+2\pi \right)$, as shown in Fig. \ref{fig12}. This result had been proposed in Ref. \cite{RN505}. Again we note that, the use of the B-K type BCs enables us to exactly solve the Fisher zeros of a finite lattice, such that the determination of the Fisher loci is rigorous.

The Fisher loci do not cut the positive real axis, thus the system in this case does not have a physical phase transition in both cases $J<0$ and $J>0$. The density function of $\left\{ \tilde z_{ij} \right\}$ on the Fisher loci had been derived \cite{RN411}. We quote it here
\begin{align}
g_{\rm{a}}\left( \theta \right) = \frac{\left(1 + \cos \theta \right)^{7/4}}{2^{5/4} \pi^2 \left| \sin \theta \right|^{5/2}} K\left[ k\left(\theta\right) \right]   \label{eq98}
\end{align}
on the arc $-\frac{1}{2} + \frac{1}{2}e^{i\theta} \left(\theta_0 \le \theta \le -\theta_0+2\pi \right)$ with
\begin{align}
&k\left(\theta \right) = \frac{1}{4} \left[\frac{3}{G\left(\theta \right)}-1\right]^{1/2} \left[1 + G\left(\theta \right) \right]^{3/2}~,   \nonumber \\
&G\left(\theta \right) = \frac{1}{\sin \frac{\theta}{2}}~,   \label{eq99}
\end{align}
and 
\begin{align}
g_{\rm{l}}\left( u \right) = \frac{-\left| 1+u \right|}{2\pi^2 u^3 k\left(u\right) H^{1/2}\left(u\right)} K\left[ \frac{1}{k\left(u\right)} \right]   \label{eq100}
\end{align}
on the line segment $-\infty < u \le -\frac{1}{2}$ with
\begin{align}
&k\left(u \right) = \frac{1}{4} \left[\frac{3}{H\left(u \right)}-1\right]^{1/2} \left[1 + H\left(u \right) \right]^{3/2}~,   \nonumber \\
&H\left(u \right) = \left(-\frac{2u+1}{u^2}\right)^{1/2}~.   \label{eq101}
\end{align}

\section{B-K type boundary conditions of the honeycomb lattice Ising model}   \label{hon}
\begin{figure} 
\includegraphics{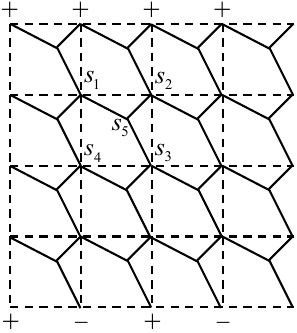}
\caption{The B-K type BCs of the honeycomb lattice Ising model.} \label{fig13}
\end{figure}

In this section we introduce the B-K type BCs for the honeycomb lattice Ising model. Like what we have done for the triangular lattice, we set a ``square'' structure in the honeycomb lattice as shown in Fig. \ref{fig13}. There are $M$ rows and $2N$ columns in the ``square'' structure, and the number of spins of the honeycomb system is $2\left(2M+1\right)N$. With the ``square'' structure the B-K type BCs are easy to determine, which are very similar to the cases of the square lattice and triangular lattice models: periodic boundary conditions in the $N$ direction, fixed spins ``$+\cdots+$'' spins in the $0$th row, and fixed spins ``$+-\cdots+-$'' in the $(M+1)$th row. Figure \ref{fig13} clearly shows the B-K type BCs.

Again we employ the mapping method as that shown in Fig. \ref{fig5}, to transform the system into an even eight-vertex model under the BCs $\left( \text{\romannumeral2} \right)$ with $M+1$ rows and $2N$ columns. Like in the cases of the square lattice and triangular lattice Ising models, the relation between the partition functions---Eq. (\ref{eq55})---holds, with the vertex weights appropriately defined using the Boltzmann factors. We note that there is a spin inside each ``square'', i.e., $s_5$ as shown in Fig. \ref{fig13}. The interactions do not consist of (or can not be represented by) the spin products associated with the ``square'' edges, such as $s_1s_2$. This indicates that the way to define the vertex weights is somewhat different from that for the square lattice and triangular lattice models.

\subsection{Zero-field case}   \label{hon-a}
In the zero field the vertex weights can be calculated from
\begin{align}
\omega = \sum\limits_{s_5 = \pm 1} e^{-\beta J s_5 \left( s_1 + s_2 + s_3 \right)}.    \label{eq102}
\end{align}
In this way the vertex weights are uniquely defined. Take vertex (1) for example. There are two possible spin states corresponding to vertex (1): $s_1 = s_2 = s_3 = s_4 = 1$ and $s_1 = s_2 = s_3 = s_4 = -1$. For either case Eq. (\ref{eq102}) produces $\omega_1 = e^{-3\beta J} + e^{3\beta J}$. Then we list all the vertex weights
\begin{align}
&\omega _1 = e^{-3\beta J} + e^{3\beta J},~ \omega _2 = \cdots = \omega_6 = e^{-\beta J} + e^{\beta J},   \nonumber \\
&\omega _7 = e^{-3\beta J} + e^{3\beta J},~ \omega _8 = e^{-\beta J} + e^{\beta J}.   \label{eq103}
\end{align}
Obviously this is a uniform free-fermion model. Using Eq. (\ref{eq34}) with the vertex weights in Eq. (\ref{eq103}), the solution turns out to be
\begin{align}
&Z_{{\rm B}\textit{-}{\rm K}, {\rm hon}} = 2^N \left[ \cosh\left(2\beta J\right) + 1 \right]^N \left[ 2\cosh\left(2\beta J\right) - 1 \right]^N   \nonumber \\
&~~~~~~~~~~~~~~\times \prod\limits_{i = 1}^N \prod\limits_{j = 1}^{M} 8 \left[ \cosh^3 \left(2\beta J\right) + 1 - \sinh^2 \left(2\beta J\right)   \right.   \nonumber \\
&~~~~~~~\left. \times \left( \cos \frac{\left( 2i - 1 \right)\pi}{2N} + 2\cos \frac{\left( 2i - 1 \right)\pi}{4N} \cos \frac{j\pi}{M+1} \right) \right].   \label{eq104} 
\end{align}
Employing the same technique as that for the triangular lattice model [see Eq. (\ref{eq79})], the partition function in the thermodynamic limit can be obtained \cite{RN122, RN123, RN221, RN224, RN558}
\begin{align}
&~~~~\mathop {\lim }\limits_{M,N \to \infty} \frac{1}{2(2M+1)N} \ln Z_{{\rm B}\textit{-}{\rm K}, {\rm hon}}   \nonumber \\
& = \frac{3}{4}\ln2 + \frac{1}{16\pi ^2}\int_0^{2\pi} d{\theta} \int_0^{2\pi} d{\phi}  \ln \left[ \cosh^3 \left(2\beta J\right) + 1  \right.   \nonumber \\ 
&~~~~~~~ \left. - \sinh^2 \left(2\beta J\right) \left( \cos{\theta} + \cos{\phi} + \cos (\theta - \phi) \right) \right].    \label{eq105}
\end{align}

\begin{figure} 
\includegraphics{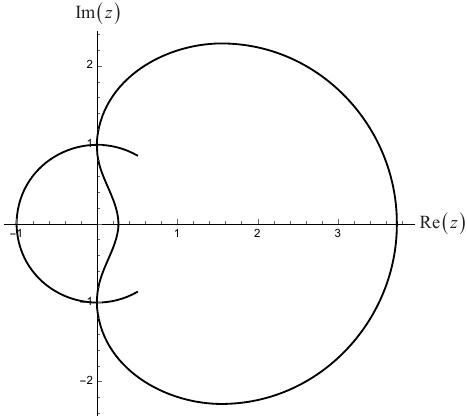}
\caption{The Fisher loci in the complex $z$ plane of the honeycomb lattice Ising model in the zero field.} \label{fig14}
\end{figure}

We express the solution in Eq. (\ref{eq104}) in terms of $z = e^{2\beta J}$
\begin{align}
&Z_{{\rm B}\textit{-}{\rm K}, {\rm hon}} = z^{-(3M+2)N} \left(z+1\right)^{2(M+1)N} \left(z^2 - z + 1\right)^N    \nonumber \\
&\times \prod\limits_{i = 1}^N \prod\limits_{j = 1}^{M} \left[ z^4 - 2z^3 + 6z^2 - 2z + 1 - 2z\left(z-1\right)^2 \left( \cos_{ij} \right) \right],    \label{eq106} 
\end{align}
where $\left( \cos_{ij} \right)$ is the same as that in Eq. (\ref{eq80}) denoting the part of the cosine functions. We see that the zeros in the complex $z$ plane include the root $-1$ of multiplicity $2(M+1)N$, the roots $\frac{1}{2} \pm \frac{\sqrt 3}{2}i$ both of multiplicity $N$, and $4MN$ roots $\left\{z_{ij} \right\}$ from the term
\begin{align}
z^4 - 2z^3 + 6z^2 - 2z + 1 - 2z\left(z-1\right)^2 \left( \cos_{ij} \right).   \label{eq107}
\end{align}
As is well-known, the interaction constant $J$ of the Ising model can be related to that of the dual lattice model, $J^*$, by \cite{RN286}
\begin{equation}
e^{2\beta J^*} = \tanh \left(-\beta J\right).   \label{eq108}
\end{equation}
We use the notation $v=\tanh \left(\beta J\right)$. Since the dual lattice of the honeycomb net is the triangular lattice, we can verify from Eq. (\ref{eq108}) that the Fisher zeros of the honeycomb lattice model in the complex $v^2$ plane coincide with those of the triangular lattice model in the complex $\tilde z$ plane (see Sec. \ref{tri-a}). To illuminate this in detail, we make use of $z=\left(1+v\right)/\left(1-v\right)$ to recast Eq. (\ref{eq107}) into 
\begin{align}
\frac{4}{\left(1-v\right)^4}\left[1 + 3v^4 - 2v^2\left(1 - v^2 \right) \left( \cos_{ij} \right) \right].   \label{eq109}
\end{align}
Comparing with Eq. (\ref{eq80}) we confirm that the Fisher loci in the complex $v^2$ plane from Eq. (\ref{eq109}) consist of the line segment $\left( -\infty, -\frac{1}{3} \right]$ and the circle $-\frac{1}{3} + \frac{2}{3}e^{i\theta} \left(0 \le \theta < 2\pi \right)$, as shown in Fig. 2 of Ref. \cite{RN458}. When transforming the Fisher loci into the complex $z$ plane, we obtain the arc $e^{i\theta} \left(\frac{\pi}{3} \le \theta \le \frac{5\pi}{3}\right)$ corresponding to the line segment $\left( -\infty, -\frac{1}{3} \right]$, and the lima-bean-shaped curve corresponding to the circle $-\frac{1}{3} + \frac{2}{3}e^{i\theta} \left(0 \le \theta < 2\pi \right)$. Figure \ref{fig14} displays the Fisher loci, which had been reported in Refs. \cite{RN436, RN457}. The analytic expression of the lima-bean-shaped curve $z=re^{i\theta}$ is derived from $\left|3v^2 + 1\right| = 2$:
\begin{align}
r = 1 \pm \sqrt{2r\cos \theta}~.   \label{eq110}
\end{align}
Ref. \cite{RN515} had proposed an equivalent expression, but for the curve in the complex $z^{-1}$ plane. We observe that Eq. (\ref{eq104}) is invariant under the transformation $J \rightarrow -J,~ z \rightarrow z^{-1}$. Therefore, the solution in terms of $z^{-1} = e^{-2\beta J}$ has the same form as Eq. (\ref{eq106}) with $z$ replaced by $z^{-1}$, and Eq. (\ref{eq110}) is also the expression in the complex $z^{-1}$ plane.

The Fisher loci cut the positive real axis at $z=2-\sqrt{3}$ and $z=2+\sqrt{3}$, which are the critical points for $J<0$ and $J>0$, respectively. The density function of $\left\{z_{ij} \right\}$ on the Fisher loci in the thermodynamic limit can be obtained from that of the variable $v^2$, i.e., Eqs. (\ref{eq85}) and (\ref{eq87}), by using $v=\left(z-1\right)/\left(z+1\right)$. The system exhibits a second-order phase transition at either critical point.

\subsection{Imaginary-field case}   \label{hon-b}
\begin{figure} 
\includegraphics{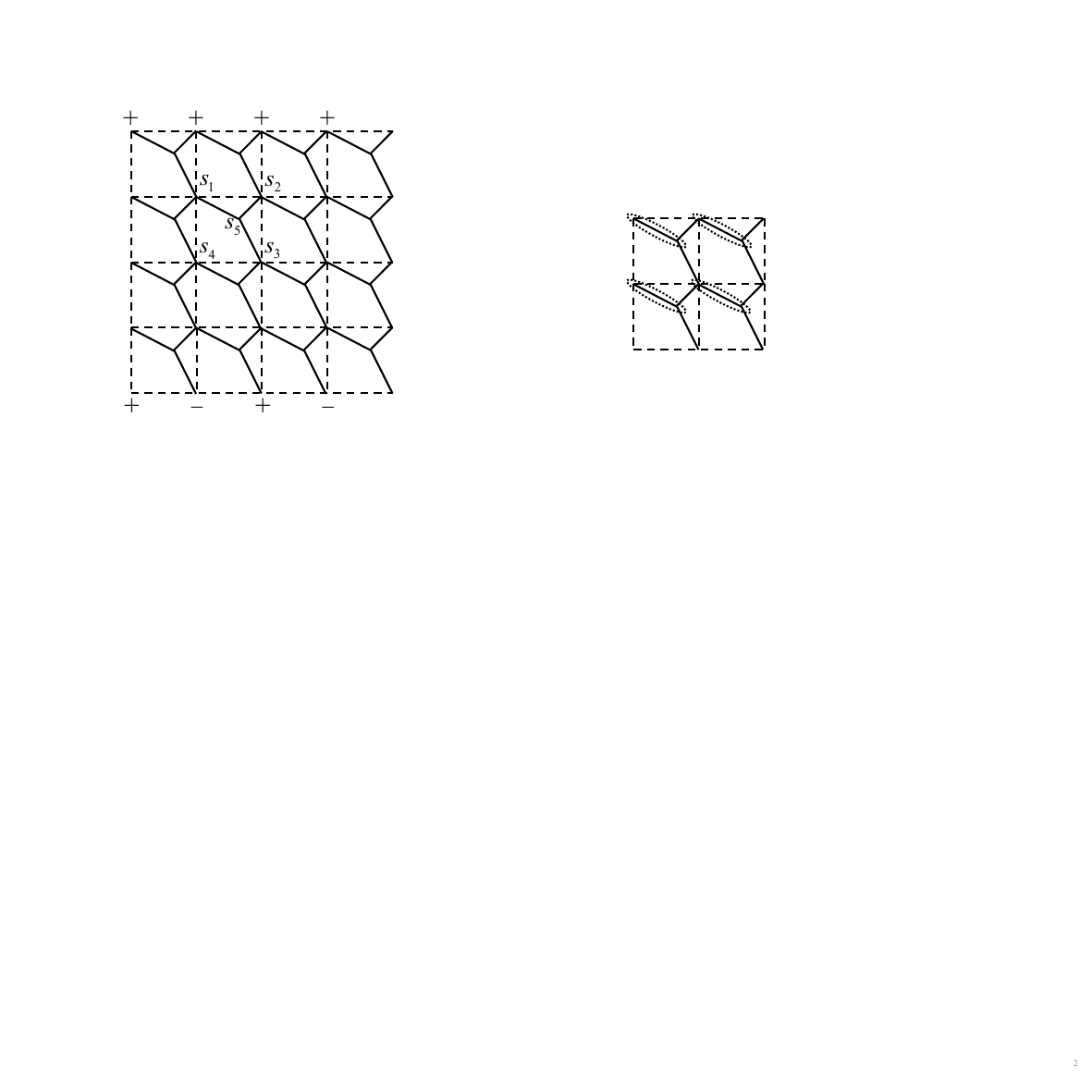}
\caption{Splitting of $\prod\nolimits_i {s_i}$ into certain pairs of product $s_i s_j$ covering all spin sites on the honeycomb lattice, with the pairs marked with dotted lines.} \label{fig15}
\end{figure}

Again, we formulate the effect of the imaginary field by the product of all spins $\prod\nolimits_i {s_i}$ according to Eq. (\ref{eq67}). Here we split $\prod\nolimits_i {s_i}$ in the way shown in Fig. \ref{fig15}, i.e., $\prod\nolimits_i {s_i}$ is split into the pairs of $s_1s_5$ in each square (see Fig. \ref{fig13}). The vertex weights are thereby determined from 
\begin{align}
\omega = \sum\limits_{s_5 = \pm 1} \left(s_1s_5\right) e^{-\beta J s_5 \left( s_1 + s_2 + s_3 \right)}.    \label{eq111}
\end{align}
We calculate all the weights
\begin{align}
&\omega _1 = \omega _7 = e^{-3\beta J} - e^{3\beta J},~\omega _3 = \omega _6 = e^{\beta J} - e^{-\beta J},    \nonumber \\
&\omega _2 = \omega _4 = \omega _5 = \omega_8 = e^{-\beta J} - e^{\beta J}.   \label{eq112}
\end{align}
This case is then mapped into a uniform free-fermion model. Substituting the vertex weights into Eq. (\ref{eq34}) yields the solution
\begin{align}
&Z_{{\rm B}\textit{-}{\rm K}, {\rm hon}, i\frac{\pi}{2}} = 2^N \left[ \cosh\left(2\beta J\right) - 1 \right]^N \left[ 2\cosh\left(2\beta J\right) + 1 \right]^N   \nonumber \\
&~~~~~~~~~~~~~~\times \prod\limits_{i = 1}^N \prod\limits_{j = 1}^{M} 8 \left[ \cosh^3 \left(2\beta J\right) - 1 - \sinh^2 \left(2\beta J\right)   \right.   \nonumber \\
&~~~~~~~\left. \times \left( \cos \frac{\left( 2i - 1 \right)\pi}{2N} + 2\sin \frac{\left( 2i - 1 \right)\pi}{4N} \cos \frac{j\pi}{M+1} \right) \right].   \label{eq113} 
\end{align}
Employing the same technique as that for the triangular lattice model [see Eq. (\ref{eq92})], the partition function in the thermodynamic limit can be obtained \cite{RN67, RN274, RN558}
\begin{align}
&~~~~\mathop {\lim }\limits_{M,N \to \infty} \frac{1}{2(2M+1)N} \ln Z_{{\rm B}\textit{-}{\rm K}, {\rm hon}, i\frac{\pi}{2}}   \nonumber \\
& = \frac{3}{4}\ln2 + \frac{1}{16\pi ^2}\int_0^{2\pi} d{\theta} \int_0^{2\pi} d{\phi}  \ln \left[ \cosh^3 \left(2\beta J\right) - 1  \right.   \nonumber \\ 
&~~~~~~~ \left. - \sinh^2 \left(2\beta J\right) \left( \cos{\phi} - \cos{\theta} + \cos (\theta+\phi) \right) \right].    \label{eq114}
\end{align}

\begin{figure} 
\includegraphics{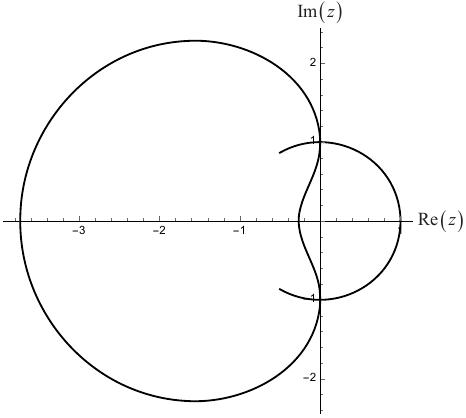}
\caption{The Fisher loci in the complex $z$ plane of the honeycomb lattice Ising model in the imaginary field.} \label{fig16}
\end{figure}

In the variable $z=e^{2\beta J}$, the solution has the expression
\begin{align}
&Z_{{\rm B}\textit{-}{\rm K}, {\rm hon}, i\frac{\pi}{2}} = z^{-(3M+2)N} \left(1-z\right)^{2(M+1)N} \left(z^2 + z + 1\right)^N    \nonumber \\
&\times \prod\limits_{i = 1}^N \prod\limits_{j = 1}^{M} \left[ z^4 + 2z^3 + 6z^2 + 2z + 1 + 2z\left(z+1\right)^2 \left( \cos_{ij} \right) \right],    \label{eq115} 
\end{align}
where $\left( \cos_{ij} \right)$ is the same as that in Eqs. (\ref{eq80}) and (\ref{eq106}). In deriving Eq. (\ref{eq115}) we have made use of
\begin{align*}
&\cos \frac{\left( 2i - 1 \right)\pi}{2N} + 2\sin \frac{\left( 2i - 1 \right)\pi}{4N} \cos \frac{j\pi}{M+1}    \nonumber \\
=& \cos \frac{\left( 2i - 1 \right)\pi}{2N} + 2\cos \frac{\left[ 2(N+1-i) - 1 \right]\pi}{4N} \cos \frac{j\pi}{M+1}     \nonumber \\
=& -\cos \frac{\left[ 2(N+1-i) - 1 \right]\pi}{2N} - 2\cos \frac{\left[ 2(N+1-i) - 1 \right]\pi}{4N}    \nonumber \\
&~~~~~~~~~~~~~~~~~~~~~~~~~~~~~~~~~~~~~\times \cos \frac{\left(M+1-j\right)\pi}{M+1}     \nonumber \\
=& -\left( \cos_{(N+1-i)(M+1-j)} \right).
\end{align*}
Comparing Eq. (\ref{eq115}) with Eq. (\ref{eq106}) we can verify that, the solution in the imaginary field is actually that in the zero field with the replacement $z \rightarrow -z$ and multiplied by $\left(-1\right)^{MN}$. This can be clearly illustrated by using the technique of defining the interaction $J^{\prime} = J - \frac{1}{\beta} \frac{\pi}{2}i$ \cite{RN70, RN274}. The Boltzmann factor associated with $J^{\prime}$ turns out to be 
\begin{align*}
e^{-\beta J^{\prime}s_1s_2} = e^{-\beta Js_1s_2} \times i s_1s_2~.
\end{align*}
This relation leads to 
\begin{align*}
Z_{{\rm B}\textit{-}{\rm K}, {\rm hon}}\left(J^{\prime}\right) = i^{6(M+1)N} \left(-1\right)^N Z_{{\rm B}\textit{-}{\rm K}, {\rm hon}, i\frac{\pi}{2}}\left(J\right).
\end{align*}
Then we have
\begin{subequations}  \label{eq116}
\begin{equation}
Z_{{\rm B}\textit{-}{\rm K}, {\rm hon}, i\frac{\pi}{2}}\left(J\right) = \left(-1\right)^{MN} Z_{{\rm B}\textit{-}{\rm K}, {\rm hon}}\left(J^{\prime}\right),    \label{eq116a}
\end{equation}
or equivalently
\begin{equation}
Z_{{\rm B}\textit{-}{\rm K}, {\rm hon}, i\frac{\pi}{2}}\left(z\right) = \left(-1\right)^{MN} Z_{{\rm B}\textit{-}{\rm K}, {\rm hon}}\left(-z\right).    \label{eq116b}
\end{equation}
\end{subequations}
The Fisher zeros in the variable $z$ in the imaginary field are exactly the additive inverses of those in the zero field. The Fisher loci in the thermodynamic limit is then easily determined, as shown in Fig. \ref{fig16} and previously reported in Ref. \cite{RN505}. The density function can also be obtained correspondingly. The Fisher loci neither cut the interval $\left(0,1 \right)$ nor $\left(1, +\infty \right)$ on the positive real axis, thus the system in this case does not exhibit a physical phase transition for both $J<0$ and $J>0$.

\section{Summary and Discussion}   \label{summary}
In this work, we report new findings in the study of statistical lattice models from two perspectives:
\begin{itemize}
\item We identify a set of BCs denoted as $\left( \text{\romannumeral2} \right)$ for the free-fermion model, under which the finite lattice partition function admits a double product form.
\item We construct the B-K type BCs for the Ising models on the triangular and honeycomb lattices, which allow the Fisher zeros of any finite lattice to be solved exactly and found to lie precisely on well-defined loci.
\end{itemize}

We formulate the Ising models within the framework of the free-fermion model. Using an appropriate mapping, the B-K and B-K type BCs are yielded from the BCs $\left( \text{\romannumeral2} \right)$. The use of B-K and B-K type BCs enables a rigorous determination of Fisher loci in the thermodynamic limit and thus is very helpful in the analysis of critical behavior. Furthermore, the product form of the partition function permits an analytical evaluation of thermodynamic quantities on a finite lattice. Therefore, the B-K and B-K type BCs are useful tools for the study of finite-size scaling and corrections \cite{RN422, RN456, RN475}. The finite-size effects of Ising models under the B-K type BCs, as well as those of the free-fermion model under the BCs $\left( \text{\romannumeral2} \right)$, merit further investigation.

It is natural to explore the construction of B-K type BCs for other two-dimensional Ising models. One notable example is the  Kagom\'e lattice model, which is not addressed in the main text. While we are able to define the B-K type BCs for the Kagom\'e lattice Ising model, we have not succeeded in expressing the corresponding partition function in a product form. Detailed discussions of this case are provided in Appendix~\ref{appb}.

\begin{acknowledgments}
This work was supported by Guangdong Provincial Quantum Science Strategic Initiative (Grants No. GDZX2203001 and No. GDZX2403001), National Natural Science Foundation of China (Grant No. 12474489), Research Funding for Outbound Postdoctoral Fellows in Shenzhen (Grant No. SZRCXM2401006), Shenzhen Fundamental Research Program (Grant No. JCYJ20240813153139050), Innovation Program for Quantum Science and Technology (Grant No. 2021ZD0302300), and Guangdong Innovative and Entrepreneurial Research Team Program (Grant No. 2016ZT06D348).
\end{acknowledgments}

\appendix
\section{Solutions of the odd free-fermion model}  \label{appa}
\begin{figure} 
\includegraphics{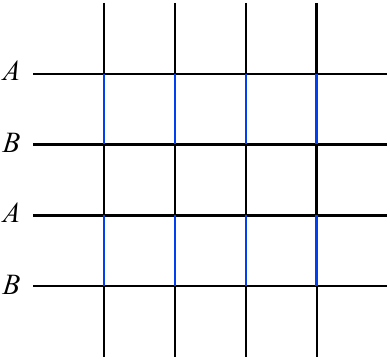}
\caption{The approach of transformation from the uniform and row-by-row staggered odd eight-vertex models to the row-by-row staggered even subcase, by reversing the arrows on the edges marked in blue.} \label{fig17}
\end{figure}

The solutions of the odd free-fermion model under the BCs $\left( \text{\romannumeral2} \right)$ can be conveniently obtained by a transformation to the even subcase, when $M$ is even. We use the approach of transformation introduced in Ref. \cite{RN129} to give the results. Figure \ref{fig17} shows this approach. By reversing the arrow directions on the vertical edges marked in blue, each configuration of the odd eight-vertex model is converted into a configuration of the even subcase, vice versa. That is, a one-to-one mapping between the configurations of the odd and even subcases is established. Then it is clear to see from Fig. \ref{fig17} that the uniform odd eight-vertex model is transformed into a row-by-row staggered even eight-vertex model, with the correspondence of the vertex weights:
\begin{subequations}  \label{eqa1}
\begin{align}
&\omega_9 \mapsto \omega_{1A},~ \omega_{10} \mapsto \omega_{2A},~ \omega_{11} \mapsto \omega_{4A},~ \omega_{12} \mapsto \omega_{3A},   \nonumber \\
&\omega_{13} \mapsto \omega_{7A},~ \omega_{14} \mapsto \omega_{8A},~ \omega_{15} \mapsto \omega_{6A},~ \omega_{16} \mapsto \omega_{5A},   \label{eqa1a} 
\end{align}
and
\begin{align}
&\omega_9 \mapsto \omega_{3B},~ \omega_{10} \mapsto \omega_{4B},~ \omega_{11} \mapsto \omega_{2B},~ \omega_{12} \mapsto \omega_{1B},   \nonumber \\
&\omega_{13} \mapsto \omega_{6B},~ \omega_{14} \mapsto \omega_{5B},~ \omega_{15} \mapsto \omega_{7B},~ \omega_{16} \mapsto \omega_{8B}.   \label{eqa1b} 
\end{align}
\end{subequations}
It is straightforward to verify that the odd free-fermion model [see Eq. (\ref{eq3})] is transformed into the row-by-row staggered even free-fermion model, and the BCs $\left( \text{\romannumeral2} \right)$ are kept under the transformation. The solution of the uniform odd free-fermion model can then be obtained via Eq. (\ref{eq52}) or Eq. (\ref{eq53}), with the replacement of weights shown in Eqs. (\ref{eqa1a}) and (\ref{eqa1b}).

The transformation in Fig. \ref{fig17} can also be applied to the row-by-row staggered odd eight-vertex model, which still produces the row-by-row staggered even subcase.  The correspondence of the vertex weights are
\begin{subequations}  \label{eqa2}
\begin{align}
&\omega_{9A} \mapsto \omega_{1A},~ \omega_{10A} \mapsto \omega_{2A},~ \omega_{11A} \mapsto \omega_{4A},~ \omega_{12A} \mapsto \omega_{3A},   \nonumber \\
&\omega_{13A} \mapsto \omega_{7A},~ \omega_{14A} \mapsto \omega_{8A},~ \omega_{15A} \mapsto \omega_{6A},~ \omega_{16A} \mapsto \omega_{5A},   \label{eqa2a} 
\end{align}
and
\begin{align}
&\omega_{9B} \mapsto \omega_{3B},~ \omega_{10B} \mapsto \omega_{4B},~ \omega_{11B} \mapsto \omega_{2B},~ \omega_{12B} \mapsto \omega_{1B},   \nonumber \\
&\omega_{13B} \mapsto \omega_{6B},~ \omega_{14B} \mapsto \omega_{5B},~ \omega_{15B} \mapsto \omega_{7B},~ \omega_{16B} \mapsto \omega_{8B}.   \label{eqa2b} 
\end{align}
\end{subequations}
We see that the row-by-row staggered odd free-fermion model ($M$ is even) is mapped into a row-by-row staggered even free-fermion model. Therefore, the solution under the BCs $\left( \text{\romannumeral2} \right)$ is still given in Eq. (\ref{eq52}) or Eq. (\ref{eq53}) when substituting the weights according to Eqs. (\ref{eqa2a}) and (\ref{eqa2b}).

We note that, the solutions of the uniform and row-by-row staggered odd free-fermion models under the BCs $\left( \text{\romannumeral2} \right)$ can also be derived by the Pfaffian method. Using the construction of the dimer lattice with the dimer weights introduced in Appendix D.1 of Ref. \cite{RN124}, one can perform the calculations similar to those we have done for the even free-fermion model. Still consider the BCs $\left( \text{\romannumeral1} \right)$ and $\left( \text{\romannumeral2} \right)$, and two Toeplitz determinants can be deduced. Again one can express the Toeplitz determinant corresponding to the BCs $\left( \text{\romannumeral2} \right)$ in a product form, thereby obtaining the solution.

\section{Discussions of the Kagom\'e lattice Ising model}  \label{appb}
\begin{figure} 
\includegraphics{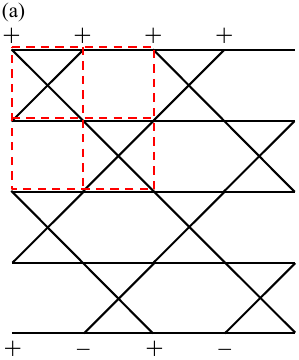} \\
~\\
\includegraphics{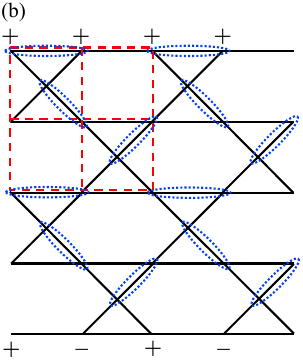}
\caption{The B-K type BCs of the Kagom\'e lattice Ising model, with the ``square'' structure marked with red dash lines. (a) The zero field case, which can be mapped into the bi-partite staggered even free-fermion model. (b) The imaginary field case, which can be mapped into the four-partite staggered even free-fermion model. $\prod\nolimits_i {s_i}$ is splitted into certain pairs of product $s_i s_j$ covering all spin sites on the Kagom\'e lattice, with the pairs marked with blue dotted lines.} \label{fig18}
\end{figure}

We can still employ the mapping method shown in Fig. \ref{fig5}, to transform the Kagom\'e lattice Ising model into the even free-fermion model. The ``square'' structure and the B-K type BCs are shown in Fig. \ref{fig18}, both for the zero field and imaginary field cases. In this way the zero field and imaginary field cases are mapped into the bi-partite and four-partite staggered even free-fermion models \cite{RN258, RN259, RN257} under the BCs $\left( \text{\romannumeral2} \right)$, respectively. However, both for the bi-partite and the four-partite staggered cases, we have not succeeded in expressing the solution under the BCs $\left( \text{\romannumeral2} \right)$ in a product form. 

In this appendix we investigate the Fisher zeros of the Kagom\'e lattice Ising model under the toroidal BCs. The boundary is defined in our recent work \cite{RN558}, as shown in Fig. 6 therein. In Ref. \cite{RN558} we have used an alternative approach to map the Ising model under the toroidal BCs into the free-fermion model, namely, the translation into the sixteen-vertex model and the weak-graph expansion for transformation into an even or odd free-fermion model. Below we quote the results of the Kagom\'e Ising model there to analyse the Fisher zeros. 

\subsection{Zero-field case}
In Ref. \cite{RN558} the zero field case is transformed into an even free-fermion model under the toroidal BCs, whose vertex weights are
\begin{align}
&\omega_1 = \frac{1}{2} \left( e^{-3\beta J} + 3e^{\beta J} \right)^2,  \nonumber  \\
&\omega_2 = \omega_3 = \omega_4 = \omega_5 = \omega_6 = \frac{1}{2} \left( e^{-3\beta J} - e^{\beta J} \right)^2,  \nonumber  \\
&\omega_7 = \omega_8 = \frac{1}{2} \left( e^{-3\beta J} + 3e^{\beta J} \right) \left( e^{-3\beta J} - e^{\beta J} \right).   \label{eqb1}
\end{align}
The finite lattice partition function of this free-fermion model is the sum of four Pfaffians (four choices of the directions on the edges of the dimer lattice, see Sec. \ref{f-f-a-1} of the present paper and Sec. IV. 5 of Ref. \cite{RN465}). As mentioned in Sec. \ref{intro}, the Onsager-Kaufman solution of the finite square lattice Ising model under the toroidal BCs \cite{RN73} consists of the sum of four products, which can actually be written as four Pfaffians. The finding of B-K BCs enables an exact calculation of the finite lattice Fisher zeros. In the case of Kagom\'e lattice Ising model, however, we have not succeeded in finding the product form of the finite lattice solution under the B-K type BCs in Fig. \ref{fig18}. Hence, we are not able to solve the Fisher zeros of a finite lattice exactly. Here we analyse the zeros of one of the four Pfaffians, and then determine the Fisher loci in the thermodynamic limit.

Consider the even free-fermion model of $M$ rows and $N$ columns, and the number of Ising spins is $3MN$. Each of the four Pfaffians is the square root of the determinant of the respective matrix. Each determinant is given in the form (one can perform the derivation like we have done in Sec. \ref{f-f} to achieve this result)
\begin{align}
\det\left({\bf{M}}\right) = \prod\limits_{i = 1}^{N} \prod\limits_{j = 1}^{M} \det \left({\bf{T}}_{ij}\right)    \label{eqb2}
\end{align}
with
\begin{align}
{\bf{T}}_{ij} = \left[ \begin{array}{*{20}{c}}
0&{\frac{\omega_8}{\omega_1}}&0&{\frac{\omega_3 - \omega _6}{\omega_1}}&{e^{i\xi_j}}&{\omega _3}\\
{ -\frac{\omega_8}{\omega_1}}&0&0&{e^{i\mu_i}}&{\frac{{{\omega _4} - {\omega _5}}}{{{\omega _1}}}}&{{\omega _4}}\\
0&0&0&1&1&{{\omega _1}}\\
{\frac{{{\omega _6} - {\omega _3}}}{{{\omega _1}}}}&{-e^{-i\mu_i}}&{ - 1}&0&{ - \frac{{{\omega _7}}}{{{\omega _1}}}}&0\\
{-e^{-i\xi_j}}&{\frac{{{\omega _5} - {\omega _4}}}{{{\omega _1}}}}&{ - 1}&{\frac{{{\omega _7}}}{{{\omega _1}}}}&0&0\\
{ -\omega_3}&{ -\omega_4}&{ -\omega_1}&0&0&0
\end{array} \right].    \label{eqb3}
\end{align}
$\mu_i$ can be $\frac{2i\pi}{N}$ or $\frac{\left(2i-1\right)\pi}{N}$, and $\xi_j$ can be $\frac{2j\pi}{M}$ or $\frac{\left(2j-1\right)\pi}{M}$. Four choices of $\left\{ \mu_i, \xi_j \right\}$ produce four determinants, thereby giving four Pfaffians. It is obvious that the four Pfaffians become degenerate in the thermodynamic limit $M,N \to \infty$. Hence, the partition function and Fisher loci in the thermodynamic limit can be determined only by one Pfaffian. We take $\mu_i = \frac{2i\pi}{N},~ \xi_j = \frac{2j\pi}{M}$. The determinant in Eq. (\ref{eqb2}) turns out to be
\begin{align}
&\det\left({\bf{M}}\right) = \prod\limits_{i = 1}^{N} \prod\limits_{j = 1}^{M} \left[ \omega_1^2 + \omega_2^2 + \omega_3^2 + \omega_4^2  \right.   \nonumber \\
&~~~~~~~ + 2\left( \omega_1 \omega_4 - \omega_2 \omega_3 \right) \cos \mu_i + 2\left( \omega_2 \omega_4 - \omega_1 \omega_3 \right) \cos \xi_j    \nonumber \\
&~~~~~~~ + 2\left( \omega_1 \omega_2 - \omega_5 \omega_6 \right) \cos (\mu_i + \xi_j)   \nonumber \\
&\left.~~~~~~ + 2\left( \omega_1 \omega_2 - \omega_7 \omega_8 \right) \cos (\mu_i - \xi_j) \right].    \label{eqb4}
\end{align}
 
With the weights in Eq. (\ref{eqb1}), we calculate this determinant for the Kagom\'e lattice Ising model
\begin{align}
&\det\left({\bf{M}}\right) = \prod\limits_{i = 1}^{N} \prod\limits_{j = 1}^{M} \left[ e^{-12\beta J} + 18e^{-4\beta J} + 24 + 21e^{4\beta J}  \right.   \nonumber \\
&\left.~~~~~~~~~ + 4\left( e^{-8\beta J} - e^{-4\beta J} - 1 + e^{4\beta J} \right) {\left( \cos_{ij} \right)}^{\prime} \right].    \label{eqb5}
\end{align}
Here the term ${\left( \cos_{ij} \right)}^{\prime}$ denotes
\begin{align*}
\cos \mu_i - \cos \xi_j + \cos(\mu_i + \xi_j)~.
\end{align*}
The well-known solution in the thermodynamic limit \cite{RN82, RN558} can be obtained by taking $\frac{1}{2} \lim_{M,N \to \infty} \frac{1}{3MN}\ln \left[ \det \left( {\bf{M}} \right) \right]$. We express this determinant in terms of $\tilde z = e^{4\beta J}$:
\begin{align}
&\det\left({\bf{M}}\right) = \tilde z^{-3MN} \prod\limits_{i = 1}^N \prod\limits_{j = 1}^M \left[ 21\tilde z^4 + 24\tilde z^3 + 18\tilde z^2 + 1 \right.   \nonumber \\
&\left.~~~~~~~~~~~~~~~~~~~~~~ + 4\tilde z \left(\tilde z + 1\right) \left(\tilde z - 1\right)^2 {\left( \cos_{ij} \right)}^{\prime} \right].    \label{eqb6} 
\end{align}
The zeros are the roots of 
\begin{align}
21\tilde z^4 + 24\tilde z^3 + 18\tilde z^2 + 1 + 4\tilde z \left(\tilde z + 1\right) \left(\tilde z - 1\right)^2 {\left( \cos_{ij} \right)}^{\prime}~.   \label{eqb7}
\end{align}
As pointed out by Ref. \cite{RN411}, the Fisher zeros of the Kagom\'e lattice model can be related to those of the honeycomb lattice model, by the star-triangle transformation and the spin decimation. The interaction constant $\tilde J$ of the honeycomb lattice model is determined by
\begin{align}
e^{-2\beta \tilde J} = \frac{e^{-4\beta J} + 1}{2}~.   \label{eqb8}
\end{align}
From this relation we can verify that
\begin{align}
\tanh^2 (\beta \tilde J) = \frac{\tanh^2 \left(2\beta J\right)}{\left( \tanh \left(2\beta J\right) + 2 \right)^2}~.   \label{eqb9}
\end{align}
Use the notations $\tilde v = \tanh \left(2\beta J\right) $ and $w = \frac{\tilde v^2}{\left(\tilde v + 2\right)^2}$. In Sec. \ref{hon-a} we have indicated that the Fisher loci of the honeycomb lattice model in the complex $\tanh^2  (\beta \tilde J)$ plane consist of the line segment $\left( -\infty, -\frac{1}{3} \right]$ and the circle $-\frac{1}{3} + \frac{2}{3}e^{i\theta} \left(0 \le \theta < 2\pi \right)$. Here we see from Eq. (\ref{eqb9}) that the Fisher loci of the Kagom\'e lattice model in the complex $w$ plane are the same. To illustrate this point, we just need to rewrite Eq. (\ref{eqb7}) as
\begin{align}
4\frac{\left(\tilde v + 2\right)^2}{\left(1 - \tilde v\right)^2} \left[1 + 3w^2 + 2w\left(1-w\right){\left( \cos_{ij} \right)}^{\prime} \right].   \label{eqb10}
\end{align}
Noticing that $-3 \le {\left( \cos_{ij} \right)}^{\prime} \le \frac{3}{2}$ and comparing Eq. (\ref{eqb10}) with Eqs. (\ref{eq109}) and (\ref{eq80}), we can confirm that the Fisher loci in the complex $w$ plane in the thermodynamic limit coincide with those of the honeycomb lattice model in the complex $\tanh^2 \left(\beta J\right)$ plane, also with those of the triangular lattice model in the complex $e^{4\beta J}$ plane. The density function in the thermodynamic limit is shown in Eqs. (\ref{eq85}) and (\ref{eq87}). 

Returning to the complex $\tilde z$ plane, the Fisher loci can be directly determined from Eq. (\ref{eqb7}) or via transformation from those in the complex $w$ plane, as shown in Fig. 5(a) of Ref. \cite{RN534}. The Fisher loci cut the positive real axis at the second-order critical point $\tilde z = \frac{2\sqrt 3}{3} - 1$. Therefore, the system exhibits a physical phase transition only when $J < 0$. The Fisher loci in other variables such as $\tanh \left(-\beta J\right)$, $e^{2\beta J}$ and $e^{-2\beta J}$ had also been reported \cite{RN458, RN534, RN457, RN437}.
 
\subsection{Imaginary-field case}
The imaginary field case of the Kagom\'e lattice Ising model is very special. In Ref. \cite{RN558} we have transformed this case under the toroidal BCs into an odd free-fermion model, with the vertex weights
\begin{align}
&\omega_9 = \omega_{11} = \omega_{13} = \omega_{15} = \frac{1}{2} e^{2\beta J} \left( 3 + e^{-4\beta J} \right) \left( 1 - e^{-4\beta J} \right),  \nonumber  \\
&\omega_{10} = \omega_{12} = \omega_{14} = \omega_{16} = -\frac{1}{2} e^{2\beta J} \left( 1 - e^{-4\beta J} \right)^2.   \label{eqb11}
\end{align}
Fortunately, the finite lattice partition function of this odd free-fermion model can be exactly solved. Still let there be $M$ rows and $N$ columns. Notice that the arrow configurations of vertices (9), (11), (13) and (15) are 1-in-3-out, while those of vertices (10), (12), (14) and (16) are 3-in-1-out (see Fig. \ref{fig1}). Under the toroidal BCs the total numbers of ``in'' and ``out'' should be equal, thus the number of each type of vertices should satisfy
\begin{align}
N_9 + N_{11} + N_{13} + N_{15} = N_{10} + N_{12} + N_{14} + N_{16}~.   \label{eqb12}
\end{align}
This requires the number of vertices $MN$ to be even and the number of spins $3MN$ to be a multiple of 6. We set $N$ to be even. Taking into account Eq. (\ref{eqb11}), the partition function is given by
\begin{align}
Z_{{\rm Kag}, i\frac{\pi}{2}} = W_{\rm con} \omega_9^{MN/2} \omega_{10}^{MN/2}.  \label{eqb13}
\end{align}
Here $W_{\rm con}$ represents the number of configurations of the $M\times N$ odd eight-vertex model. 

$W_{\rm con}$ is easy to calculate. Consider the top row of vertices with $N$ upper vertical arrows and $N$ lower vertical arrows. From Fig. \ref{fig1} we can see that, if the upper and lower vertical arrows of a vertex are in the same direction (either both $\uparrow$ or both $\downarrow$), the two horizontal arrows of this site are pointing in the opposite direction (either $\leftarrow \rightarrow$ or $\rightarrow \leftarrow$). When the upper and lower vertical arrows are in the opposite direction, the two horizontal arrows keep the direction unchanged. Due to the periodic boundary conditions along the $N$ direction, there should be an even number of lower vertical arrows which point in the same direction as the respective upper vertical arrows (the sum of the numbers of $\leftarrow \rightarrow$ sites and of $\rightarrow \leftarrow$ sites should be even). Hence, given a set of upper vertical arrows, the number of choices for lower vertical arrows is $C^0_N + C^2_N + \cdots + C^N_N = 2^{N-1}$. With fixed upper and lower vertical arrows there are two choices for the horizontal arrows (one is the inverse version of the other, i.e., all horizontal arrows are reversed), thus the number of the resulting configurations of this row is $2 \times 2^{N-1} = 2^N$. We can repeat this process from the top row to the $\left(M-1\right)$th row, and take into account the $2^N$ initial choices for the top row of upper vertical arrows (which are also the bottom row of lower vertical arrows). The number of configurations turns out to be
\begin{align}
W_{\rm con} = \left(2^N\right)^{M-1} \times 2 \times 2^N = 2^{MN+1}~.   \label{eqb14}
\end{align}
The partition function of this case can then be determined from Eq. (\ref{eqb13})
\begin{align}
Z_{{\rm Kag}, i\frac{\pi}{2}} = 2 \tilde z^{-3MN/2} \left[ -\left(3\tilde z + 1\right) \left(\tilde z - 1\right)^3 \right]^{MN/2},  \label{eqb15}
\end{align}
where $\tilde z = e^{4\beta J}$. The solution is surprisingly simple. It is straightforward to give the partition function in the thermodynamic limit, as shown in Ref. \cite{RN558}. Unlike most of the two-dimensional Ising models, in this case the Fisher zeros of any finite lattice are located at two points in the complex $\tilde z$ plane: $\tilde z = -\frac{1}{3}$ and $\tilde z = 1$. The Fisher loci do not form continuous curves. It is obvious that the system has not a physical phase transition. The solution of this case demonstrates ``the special role played by the Kagom\'e lattice'', which is quoted from Ref. \cite{RN144}.

\end{document}